\documentclass[review]{elsarticle}

\usepackage{lineno,hyperref}
\usepackage{changepage}
\usepackage{geometry}
\usepackage{amsmath}
\usepackage{multirow}
\usepackage{multicol}
\modulolinenumbers[1]
\usepackage{changes}
\usepackage{lineno}
\usepackage{color}
\usepackage{CJK}
\usepackage{enumitem}
\usepackage{graphicx}
 \usepackage{algpseudocode}
\usepackage{algorithm}
\usepackage{upgreek}
\usepackage{lineno}

\journal{Journal of Physics: Condensed Matter}

\begin{document}

\begin{frontmatter}

\title{GiftBTE: An efficient deterministic solver for non-gray phonon Boltzmann transport equation}

\author{Yue Hu$^{1,2,\#}$, Ru Jia$^{1,2,\#}$, Jiaxuan Xu$^{1,2,\#}$, Yufei Sheng$^{1,2}$, Minhua Wen$^{3}$, James Lin$^{3}$, Yongxing Shen$^{2}$, Hua Bao$^{1,2,*}$}
\address{$^1$Global Institute of Future Technology, Shanghai Jiao Tong University, Shanghai 200240, P. R. China}

\address{$^2$University of Michigan–Shanghai Jiao Tong University Joint Institute, Shanghai Jiao Tong University, Shanghai 200240, P. R. China}

\address{$^3$Center for High Performance Computing, Shanghai Jiao Tong University, Shanghai 200240, P. R. China}
\address{$^\#$These authors contributed equally to this work}
\address{*Corresponding author: Email: hua.bao@sjtu.edu.cn}

\begin{abstract}
Advances in nanotechnology have facilitated the exploration of submicron thermal transport. At this scale, Fourier's law is no longer applicable, and the governing equation for thermal transport is the phonon Boltzmann transport equation (BTE). However, the availability of open-source solvers for the phonon BTE is limited, impeding progress in this field. This study introduces an open-source package, GiftBTE, for numerically solving the non-gray phonon BTE. GiftBTE employs deterministic solutions and provides both steady-state and transient solvers. For the steady-state solver, GiftBTE employs the implicit discrete ordinates method (DOM) with second-order spatial accuracy and the synthetic iterative scheme. For the transient solver, GiftBTE employs the explicit DOM with second-order spatial accuracy. This package demonstrates excellent computational efficiency, enabling realistic three-dimensional simulations of devices and materials. By interfacing with first-principles calculations, this solver enables parameter-free computation of submicron thermal transport. The application of GiftBTE includes, but is not limited to, computing the thermal conductivity of nanostructures, predicting temperature rises in transistors, and simulating laser heating processes.
\end{abstract}




\end{frontmatter}


\section{Introduction} \label{sec:intro}
Advances in nanotechnology have facilitated the exploration of thermal transport at the micro/nanoscale. For example, commercial silicon fin field-effect transistors (FinFET) \citep{sinha2012exploring} have a characteristic size of ten nanometers. Even more compact devices, like gate-all-around or two-dimensional devices, have also been proposed and studied \citep{chhowalla2016two,liu2020two}. With the aggressive downsizing of these devices, the energy density becomes quite substantial, and heat conduction is the sole means of heat dissipation on this small scale, becoming a significant bottleneck that hinders further development of electronics \citep{moore2014emerging}. The fabrication of nanometer-sized structures in materials also becomes crucial in reducing the thermal conductivity for applications of thermal barriers or thermoelectric conversion \citep{venkatasubramanian2001thin,hochbaum2008enhanced}. Therefore, micro/nanoscale thermal transport has become a major topic of interest, attracting significant research attention \citep{cahill2003nanoscale,chen2021non}.

At the micro/nanoscale, it has been well established that Fourier's law is no longer valid \citep{casimir1938note}. Several methods for simulating thermal transport at these scales have been developed \citep{majumdar1993microscale,volz1999molecular,broido2007intrinsic,bao2018review}, including lattice dynamics \citep{broido2007intrinsic,turney2009predicting}, non-equilibrium Green's function \citep{mingo2003phonon,wang2008quantum}, molecular dynamics (MD) \citep{mcgaughey2006phonon,volz1999molecular}, and the phonon Boltzmann transport equation (BTE) \citep{majumdar1993microscale,klitsner1988phonon}. While the key theoretical developments for these methods have solidified our understanding of micro/nanoscale thermal transport \citep{majumdar1993microscale,mazumder2022boltzmann}, the release of open-source packages for some of these methods has undoubtedly aided the rapid progress of this field \citep{lindsay2018survey,gu2021thermal}. For example, there are several open-source packages for lattice dynamics, such as ShengBTE \citep{li2014shengbte}, AlmaBTE \citep{carrete2017almabte}, phono3py \citep{togo2015distributions}, and ALAMODE \citep{tadano2014anharmonic}. For MD simulations, LAMMPS \citep{plimpton1995fast} and GPUMD \citep{fan2017efficient} have been developed. Using these packages, a substantial portion of research has explored thermal transport in atomic systems, including size-dependent thermal conductivity \citep{shiomi2006non,luo2012enhancement}, heat conduction in low-dimensional materials \citep{nika2017phonons,huang2018thermal}, and interfacial thermal resistance \citep{shao2015molecular,chen2022interfacial}. However, lattice dynamics, non-equilibrium Green's function, and MD are generally only applicable at the atomic scale, i.e., less than 10 nm, due to computational constraints \citep{fan2017efficient,guo2022atomistic}. Since atomic-scale structures are usually not directly applicable in practical applications, they must be assembled into larger systems \citep{chen2014multiscale,hu2021perspective}. The computational cost of atomic-scale methods is prohibitive for simulating these large systems \citep{fan2017efficient}, and the macroscopic heat diffusion equation is invalid due to the local nanostructures \citep{bao2018review,chen2021non}. Therefore, the submicron method, i.e., phonon BTE, which bridges the atomic and macroscopic scales, should be adopted.

However, in comparison to the robust open-source packages for lattice dynamics and MD, the available open-source solvers for the phonon BTE are much less. Some lattice dynamics tools, such as ShengBTE \citep{li2014shengbte}, AlmaBTE \citep{carrete2017almabte}, phono3py \citep{togo2015distributions}, and ALAMODE \citep{tadano2014anharmonic}, involve solving the phonon BTE, but are only suitable for bulk systems and simple geometries, such as thin films and nanowires. 
Recently, several open-source packages that are based on the numerical solution of the phonon BTE have been developed. These packages include OpenBTE \citep{romano2021openbte}, Ptrans \citep{shao2022p}, and MCBTE \citep{pathak2021mcbte}. The development of these packages has been driven by the increasing maturity of numerical methods and the availability of first-principles phonon properties. These packages can handle more complex geometries, but are still restricted in their ability to compute thermal conductivity for typical nanostructures, such as nano-porous media, thin films, and nano-composites. These open-source packages cannot address other important applications of the phonon BTE, such as predicting the temperature rise in transistors and simulating laser heating processes \citep{mazumder2022boltzmann}. As such, there is still a lack of a general-purpose package that can function like conventional CAE software.

In this study, we introduce GiftBTE, an open-source general-purpose package for the phonon BTE. GiftBTE utilizes a deterministic approach to solve the phonon BTE, which overcomes statistical errors and achieves higher accuracy compared to packages that employ Monte Carlo methods\citep{pathak2021mcbte, shao2022p}. Unlike OpenBTE \citep{romano2021openbte}, a previous deterministic package that offers a steady-state solver based on the implicit discrete ordinates method (DOM) with first-order spatial accuracy, the steady state solver in GiftBTE provides the implicit DOM with second-order spatial accuracy and synthetic iterative schemes. The second-order spatial accuracy can largely reduce the required number of meshes in accurate numerical solutions \cite{hu2022ultra} and the synthetic iterative schemes can overcome the slow convergence rate in near-diffusive regime  \citep{zhang2021fast}. This solver exhibits significantly improved computational efficiency and enables simulations of realistic three-dimensional devices and materials. Moreover, GiftBTE offers solutions for the phonon BTE with a heat source and the transient phonon BTE, enabling predictions of temperature rise in transistors and simulations of laser heating processes.

This manuscript provides a comprehensive description of the GiftBTE. It is organized as follows. In Section \ref{sec:BTE}, we present the governing equation and boundary conditions of the phonon BTE. In Section \ref{sec:steady}, we introduce the steady state solver of the phonon BTE within the framework of the GiftBTE. In Section \ref{sec:transient}, we provide the transient solver of the phonon BTE in the GiftBTE. Section \ref{sec:workflow} details the entire workflow of GiftBTE. In Section \ref{sec:investigation}, we investigate the heat conduction problems in nanostructures, transistors and laser heating processes using the GiftBTE. Finally, in Section \ref{sec:conclu}, we draw conclusions.

\section{Phonon BTE} \label{sec:BTE}

\subsection{Governing equation}
The phonon BTE describes the transport of phonons and is expressed as follows \citep{ziman1960electrons}:
\begin{equation}
    \frac{{\partial f}}{{\partial t}} + {\bf{v}} \cdot \nabla f = {\left[ {\frac{{\partial f}}{{\partial t}}} \right]_{{\rm{scat}}}},\label{eq:BTE_f_scatter}
\end{equation} 
where $f (\omega, p, \bf{s}, \bf{r})$ is the distribution function, which relates to frequency $\omega$, branch $p$, propagation direction $\bf{s}$, and location $\bf{r}$. $\bf{v}$ is the group velocity. The left-hand side illustrates the change in distribution function resulting from the unencumbered flight of phonons, whereas the right-hand side accounts for the change in distribution function resulting from phonon scattering \citep{ kittel1986introduction}. As the primary mechanism of thermal transport in semiconductors is phonon transport, the BTE for phonons effectively describes the thermal transport process \citep{ kittel1986introduction,ziman1960electrons}. However, the scattering term in the phonon BTE is quite intricate, and therefore, to simplify the calculation of its practical applications, the relaxation time approximation (RTA) is commonly employed \citep{bhatnagar1954model, ziman1960electrons, kittel1986introduction}. In the context of the RTA, the phonon BTE is expressed as \citep{bhatnagar1954model,ziman1960electrons, majumdar1993microscale}:

\begin{equation}
    \frac{{\partial f}}{{\partial t}} + {\bf{v}} \cdot \nabla f =  - \frac{{f - {f^{\rm{0}}}}}{\tau } + s.\label{eq:BTE_f_RTA}
\end{equation}
Within the current context, the scattering mechanisms, including phonon-phonon scattering and phonon-defect scattering, are integrated into a unified relaxation process. The relaxation time, represented by $\tau$, quantifies the rate of decay of thermal equilibrium deviations. Meanwhile, the scattering mechanisms involving other types of carriers, such as electrons, are incorporated into the source term $s$. Previous packages commonly ignored the source term for the phonon BTE. With this simplification, one cannot predict the temperature field for transistors, in which the heat source is generated by electron-phonon interactions \citep{pop2010energy}. In this package, the source term is considered. 

 By defining the phonon energy density distribution as $e = \hbar \omega D\left( {\omega ,p} \right)(f - {f^0}\left( {{T_{{\rm{ref}}}}} \right))$, the energy-based phonon BTE can be expressed as \citep{loy2013fast}:

\begin{equation}
 \frac{{\partial e}}{{\partial t}} + {\bf{v}} \cdot \nabla e =  - \frac{{e - {e^0}}}{\tau } + \dot Q, \label{eq:BTE_origin}
\end{equation}
where ${e^{{\rm{0}}}}$ is the energy density distribution function at equilibrium state, defined as ${e^0} = \hbar \omega D\left( {\omega ,p} \right)({f^0} - {f^0}\left( {{T_{{\rm{ref}}}}} \right))$. $T_{\rm ref}$ is the reference temperature. The volumetric heat generation term, $\dot Q$, can arise from the movement of electrons through electron-phonon interactions \citep{pop2005monte,joseph2022electron,miao2021nonequilibrium}.

According to the energy conservation \citep{bhatnagar1954model,ziman1960electrons, kittel1986introduction}, ${e^0}$ is related to $e$ through

\begin{equation}
\begin{array}{{c}}
{{e^0} = \frac{1}{{4\pi }}C{T_L},}\\
{{T_L} = \frac{{\int\limits_{4\pi } {\sum\limits_p {\int_{{\omega _{\min }}}^{{\omega _{\max }}} {\frac{e}{\tau }d\omega d\Omega } } } }}{{\sum\limits_p {\int_{{\omega _{\min }}}^{{\omega _{\max }}} {\frac{C}{\tau }d\omega } } }},}
\end{array} \label{eq:BTE_get_e0}
\end{equation}
where $C$ is the volumetric heat capacity.  $T_L$ is the lattice temperature \citep{hao2009frequency}. $\Omega$ is the solid angle in spherical coordinates. Note that this expression is for three-dimensional materials. For two-dimensional materials, $\Omega$ should be replaced with $\theta$, where $\theta$ is the polar angle in the polar coordinates system. In the following discussion, we present the formation of three-dimensional materials for simplicity. In the GiftBTE package, we also consider two-dimensional materials. Equations (\ref{eq:BTE_origin}) and (\ref{eq:BTE_get_e0}) are the closed form of the phonon BTE. If proper boundary conditions are specified, they can be numerically solved to obtain the local energy density $e$. The lattice temperature is defined to ensure the energy conversion of the scattering process, which is different from the definition of local temperature $T$ \cite{ran2019efficiency,chiloyan2021green}:
\begin{equation}
   T = \frac{{\int\limits_{4\pi } {\sum\limits_p {\int_{{\omega _{\min }}}^{{\omega _{\max }}} {ed\omega d\Omega } } } }}{{\sum\limits_p {\int_{{\omega _{\min }}}^{{\omega _{\max }}} {Cd\omega } } }}.   
\end{equation}
The heat flux is calculated by
\begin{equation}
{\bf{q}} = \int\limits_{4\pi } {\sum\limits_p {\int_{{\omega _{\min }}}^{{\omega _{\max }}} {{\bf{v}}ed\omega d\Omega } } }  . 
\label{eq:BTE_get_q}
\end{equation}

All phonon properties, including the heat capacity ($C$), relaxation time ($\tau$), and group velocity ($\bf{v}$) serve as inputs for the solver. These properties can be obtained from first-principles calculations \cite{li2014shengbte,lindsay2018survey,noffsinger2010epw}. GiftBTE provides the interface with two first-principles calculations packages \cite{tadano2014anharmonic,li2014shengbte}, which will be discussed in Sec. \ref{sec:workflow}.

\subsection{Boundary condition}
\label{sec:BC}

In the context of phonon BTE, there are five commonly employed boundary conditions. 

(i) At thermalizing boundaries, all phonons are emitted at an equilibrium state of a temperature of ${T_1}$, as described by the following equation:

\begin{equation}
e = {\rm{  }}\frac{C}{{4\pi }}{T_1}\left( {{\bf{s}} \cdot {\bf{n}} < 0} \right), \label{eq:thermal}
\end{equation}
where $\bf{n}$ denotes the exterior normal unit vector of the boundary. This type of boundary is typically found at the interface between metals and semiconductors, as well as at the boundary that lies far away from the ballistic regime \citep{majumdar1993microscale}.

(ii) The specularly reflecting boundary condition is an adiabatic boundary condition in which the reflected angle of the phonon is equal to the incident angle, as stated by \citep{ziman1960electrons}:

\begin{equation}
e({\bf{s}}) = e({\bf{s}'})({\bf{s}} \cdot {\bf{n}} < {\bf{0}}),
\end{equation}
where $\bf{s}' =\bf{s}- \bf{s}(\bf{s}\cdot {\bf{n}})$ is the incident direction before being reflected to direction ${{\bf{s}}}$. This boundary is symmetric, as it cuts the symmetry domain into half \citep{hu2020unification,sheng2022size}.

(iii) The diffusely reflecting boundary condition is another adiabatic boundary condition in which the energy of the reflected phonon is the same along each direction, as defined by \citep{ziman1960electrons}:

\begin{equation}
e({\bf{s}}) = \frac{1}{\pi }\int_{{{\bf{s}}^\prime } \cdot {\bf{n}} > 0} {e({{\bf{s}}^\prime })} {{\bf{s}}^\prime } \cdot {\bf{n}}d\Omega {\rm{ (}}{\bf{s}} \cdot {\bf{n}} < 0).
\label{eq:diff_bound}
\end{equation}
This boundary condition exists at a rough surface or a surface with an amorphous oxide layer \citep{shao2018understanding,racichandran2018spectrally}.

(iv) The periodic boundary condition represents two boundaries that are connected to each other, as described by:

\begin{equation}
e({{\bf{r}}_{B1}}) = e({{\bf{r}}_{B2}})\left( {{\bf{s}} \cdot {\bf{n}} < 0} \right),\label{eq:periodic}
\end{equation}
where B1 and B2 are indexes of the two periodic boundaries. In some cases, a temperature difference $\Delta T$ is applied between the two periodic boundaries to simulate a uniform temperature gradient, leading to the equation:

\[e({{\bf{r}}_{B1}}) = e({{\bf{r}}_{B2}}) + \frac{1}{{4\pi }}C\Delta T.\]

(v) The interfacial boundary condition represents that some phonons are diffusely reflected by the interface, and some phonons diffusely transmit across the interface, i.e., \cite{ran2019efficiency,singh2009effect}
\begin{equation}
\begin{aligned}
e({\bf{s}},{{\bf{r}}_A}) = \frac{{{\beta _{BA}}}}{\pi }\int\limits_{{{\bf{s}}^\prime } \cdot {\bf{n}} < 0} {e({{\bf{s}}^\prime },{{\bf{r}}_B}){{\bf{s}}^\prime } \cdot {\bf{n}}d\Omega }  + \frac{{1 - {\beta _{AB}}}}{\pi }\int\limits_{{{\bf{s}}^\prime } \cdot {\bf{n}} > 0} {e({{\bf{s}}^\prime },{{\bf{r}}_A}){{\bf{s}}^\prime } \cdot {\bf{n}}d\Omega }  \left( {{\bf{s}} \cdot {\bf{n}} < 0} \right), 
\end{aligned}
\end{equation}
where the indexes $A$ and $B$ represent the material on two sides of the interface. The ${\beta _{{{BA}}}}$ represents the percentage of energy transmitting from B to A, and ${\beta _{{{AB}}}}$ represents the percentage of energy transmitting from A to B. The ${\beta _{{{BA}}}}$ usually equals to $1-{\beta _{{{AB}}}}$, which can be calculated by the spectral diffusive mismatch model \cite{ran2019efficiency,singh2009effect}. The transmittance $\beta$ also serves as the input of the solver when employing the interfacial boundary condition. 

These five types of boundary conditions are supported by GiftBTE and can be defined by users in input data, which will be discussed in Sec. \ref{sec:workflow}.

\section{Steady state solver} \label{sec:steady}

At steady state, the equation (\ref{eq:BTE_origin}) is expressed as: 
\begin{equation}
{\bf{v}} \cdot \nabla e =  - \frac{{e - {e^0}}}{\tau } + \dot Q. 
\label{eq:steady}
\end{equation}
$e^{{\rm{0}}}$ is related to $e$ through

\begin{equation}
\begin{array}{{c}}
{{e^0} = \frac{1}{{4\pi }}C{T_L},}\\
{{T_L} = \frac{{\int\limits_{4\pi } {\sum\limits_p {\int_{{\omega _{\min }}}^{{\omega _{\max }}} {\frac{e}{\tau }d\omega d\Omega } } } }}{{\sum\limits_p {\int_{{\omega _{\min }}}^{{\omega _{\max }}} {\frac{C}{\tau }d\omega } } }}.}
\end{array} 
\end{equation}

In GiftBTE, this equation is solved by the implicit DOM. To numerically solve this equation using the DOM, one first needs band discretization, directional discretization, and spatial discretization to transform the integro-differential equation into algebraic equations. The algebraic equations are then solved iteratively. The major difference between the GiftBTE and the previous deterministic package \cite{romano2021openbte} is that the GiftBTE achieves spatial discretization with second-order accuracy (the previous package adopts spatial discretization with first-order accuracy) and provides a synthetic iterative scheme to iteratively solve the algebraic equations. The second-order spatial accuracy can significantly reduce the required number of meshes in an accurate numerical solution \cite{hu2022ultra} and the synthetic iterative schemes can overcome the slow convergence rate in the near-diffusive regime \cite{zhang2021fast}.

\subsection{Band discretization}

 The band discretization transforms integrations associated with frequency $\omega$ into summations:
 \begin{equation}
\begin{array}{c}
\sum\limits_p {\int_{{\omega _{\min }}}^{{\omega _{\max }}} {\frac{e}{\tau }d\omega } }  = \sum\limits_\lambda  {\frac{{{e_\lambda }}}{{{\tau _\lambda }}}}  ,\\
\sum\limits_p {\int_{{\omega _{\min }}}^{{\omega _{\max }}} {\frac{C}{\tau }d\omega } }  = \sum\limits_\lambda  {\frac{{{C_\lambda }}}{{{\tau _\lambda }}}} .
\end{array}\\ 
\end{equation}
The sampled phonon band $\lambda$ and its properties ( heat capacity ($C$), relaxation time ($\tau$), and group velocity ($\bf{v}$)) are determined by the mean free path domain discretization scheme \cite{hu2022ultra}. We first collect all input phonon properties (including heat capacity $C$, group velocity ${\bf{v}}$, and relaxation time $\tau$ from lattice dynamics calculations) and find the maximum mean free path $\Lambda_{max}$ and minimum mean free path $\Lambda_{min}$ ($\Lambda=v\tau$). Then we divide the mean free path domain $[\Lambda_{min},\Lambda_{max}]$ into several bands $[\Lambda_0,..,\Lambda_n]$. For each band, we obtain the representative phonon properties as: 

 \begin{equation}
\begin{array}{c}
{C_\lambda } = \sum\limits_{v\tau  >  = {\Lambda _{n - 1}}}^{{\Lambda _n}} C ,\\
{v_\lambda } = \frac{{\sum\limits_{v\tau  >  = {\Lambda _{n - 1}}}^{{\Lambda _n}} {Cv} }}{{\sum\limits_{v\tau  >  = {\Lambda _{n - 1}}}^{{\Lambda _n}} C }},\\
{\tau _\lambda } = \frac{{\sum\limits_{v\tau  >  = {\Lambda _{n - 1}}}^{{\Lambda _n}} {C{v^2}\tau } }}{{{v_\lambda }\sum\limits_{v\tau  >  = {\Lambda _{n - 1}}}^{{\Lambda _n}} {Cv} }},
\end{array} \label{eq:banddis}
\end{equation}
where $\lambda$ denotes the index of the sampled phonon band.  In
this discretization, we collect all input phonon properties and assumed that phonon properties are isotropic (independent of velocity direction). This assumption is widely
adopted in previous studies \citep{yang2013mean,murthy2003improved}. It is valid for many important materials such as Si, GaN, etc.
The number of the sampled band $\lambda$ also serves as the input of GiftBTE package, which will be discussed in Sec. \ref{sec:workflow}. 

\subsection{Directional discretization}
There are also several integrations over the control angles $\Omega$. Directional discretization transforms those integrations into summations. For example,

\begin{equation}
\begin{array}{c}
\int\limits_{4\pi } {{e_\lambda }d\Omega }  = \int\limits_0^{\pi } {\int\limits_0^{2\pi}  {{e_\lambda }\sin \theta d\theta d\varphi } }  = \sum\limits_\alpha  {{w_{{\theta _\alpha }}}{w_{{\varphi _\alpha }}}\sin {\theta _\alpha }{e_{\alpha ,\lambda }}}  = \sum\limits_\alpha  {{w_\alpha }{e_{\alpha ,\lambda }}}. 
\end{array}
\label{eq:directiondis}
\end{equation}
where $\theta$ is the polar angle and $\varphi$ is the azimuth angle. $\alpha$ is the index of the sampled direction. $w_{{\theta _\alpha }}$ and $w_{{\varphi _\alpha }}$ are the weights of the sampled polar angle $\theta_\alpha$ and azimuth angle $\varphi_\alpha$. The sampled direction is thus $\bf{s}_{\alpha}$ $=\{\sin \theta_\alpha \cos \varphi_\alpha$ $, \sin \theta_\alpha \sin \varphi_\alpha, \cos \theta_\alpha\}$. The total weight of this direction is $w_\alpha=w_{{\theta _\alpha }}w_{{\varphi _\alpha }}\sin {\theta _\alpha }$. The weights and the corresponding moving directions are obtained by the Gauss-Legendre quadrature over the first quadrant $[\theta_{min}=0,\theta_{max}=\pi/2]$ and $[\varphi_{min}=0,\varphi_{max}=\pi/2]$. The weights and the corresponding moving directions for other quadrants are obtained according to symmetry \cite{guo2016discrete}. The total number of sampled directions $\bf{s}$ is the number of polar angles $\theta$ $\times$ the number of azimuth angles $\varphi$ $\times$ 8. The numbers of $\theta$ and $\varphi$ also serve as the input of the GiftBTE package, which will be discussed in Sec. \ref{sec:workflow}.

\subsection{Spatial discretization}

By adopting band discretization and directional discretization, Equation (\ref{eq:steady}) becomes a partial differential equation:

\begin{equation}
{{\bf{v}}_\lambda } \cdot \nabla {e_{\alpha ,\lambda }} =  - \frac{{{e_{\alpha ,\lambda }} - e_\lambda ^{\rm{0}}}}{{{\tau _\lambda }}} + {\dot Q_\lambda }, 
\end{equation}
where the equilibrium energy density can be obtained by:
\begin{equation}
\begin{array}{*{20}{c}}
{e_\lambda ^0 = \frac{1}{{4\pi }}{C_\lambda }{T_L},}\\
{{T_L} = \frac{{\sum\limits_\lambda  {\sum\limits_\alpha  {{w_\alpha }\frac{{{e_{\alpha, \lambda}}}}{{{\tau _\lambda }}}} } }}{{\sum\limits_\lambda  {\frac{{{C_\lambda }}}{{{\tau _\lambda }}}} }}.}
\end{array} \label{eq:BTE_get_e0}
 \end{equation}

Spatial discretization further transforms the partial differential equation into algebraic equations. GiftBTE adopts the finite volume method with the delta-form to discrete the equation: 

\begin{equation}
\begin{array}{l}
\frac{{\Delta e_{i,\alpha ,\lambda }^{n + 1}}}{{{\tau _{i,\lambda }}}} + \frac{1}{{{V_i}}}\sum\limits_{j \in N\left( i \right)} {\Delta e_{ij,\alpha ,\lambda }^{n + 1}{{\bf{v}}_{i,\lambda }} \cdot {{\bf{n}}_{ij}}{S_{ij}}}  = \\
 - \frac{1}{{{V_i}}}\sum\limits_{j \in N\left( i \right)} {e_{ij,\alpha ,\lambda }^n{{\bf{v}}_{i,\lambda }} \cdot {{\bf{n}}_{ij}}{S_{ij}}}  - \frac{{e_{i,\alpha ,\lambda }^n - e_{i,\lambda }^{n,{\rm{0}}}}}{{{\tau _{i,\lambda }}}} + {{\dot Q}_{i,\lambda }}
\end{array}
\label{eq:disretized_BTE}
\end{equation}
where $\Delta {e^{n + 1}} = {e^{n + 1}} - {e^n}$ and $n$ is the iteration index. ${V_i}$ is the volume of the spatial cell $i$, $N\left( i \right)$ is the set of face neighboring cells of cell $i$, $ij$ is the face between cell $i$ and cell $j$, ${S_{ij}}$ is the area of the face $ij$, and ${{\bf{n}}_{ij}}$ is the exterior normal unit vector of the face $ij$ directing from cell $i$ to cell $j$. When the system is at steady state, the left-hand side is zero and only the right-hand side remains. The first-order up-wind scheme is adopted for the $\Delta {e_{ij}}$ to ensure the sparsity and the second-order scheme is adopted for the ${e_{ij}}$ to ensure the piece-wise linear results, which is crucial for eliminating the limitation of mesh size \citep{luo2017discrete,zhang2021fast}. For the first-order up-wind scheme, the energy on the face is the energy density of the cell in the up-wind direction, i.e., ${\bf{s}} \cdot {\bf{n}} > 0$ for this cell. For the second-order scheme, the energy density ${e_{ij,\bf{s}}}$ on the face is calculated from the cell-centered energy density and the gradient of the energy density by
\begin{equation}
{e_{ij}}={e_m} + \nabla {e_m} \cdot {{\bf{l}}_{m,ij}},
\end{equation}
where cell $m$ is the cell in the upwind direction, and ${{\bf{l}}_{m,ij}}$ is the vector from the center of cell $m$ to the center of the face $ij$. The gradient $\nabla {e}$ is calculated by the least squares method, which performs well for unstructured meshes. The meshes also serve as inputs of the GiftBTE, which will be discussed in Sec. \ref{sec:workflow}.

\subsection{Sequential iterative scheme and synthetic iterative scheme}
The total number of the algebraic equations (Eq. \ref{eq:disretized_BTE}) is the number of bands $N_\lambda$ $\times$ the number of directions $N_\alpha$ $\times$ the number of cells $N_i$, which can be extremely large \cite{ali2014large}. Directly solving the linear system for all these algebraic equations is impractical. Therefore, the iterative method is commonly adopted. GiftBTE provides two iterative schemes. One is the sequential iterative scheme \cite{loy2013fast} and the other is the synthetic iterative scheme \cite{zhang2021fast}. 

In the sequential iterative scheme, the initial equilibrium energy density is guessed and the linear systems for all bands $\lambda$ and directions $\bf{s}_\alpha$ are solved sequentially. Within each iteration, the equilibrium energy density is computed according to the equation (\ref{eq:BTE_get_e0}) after solving all bands $\lambda$ and directions $\bf{s}_\alpha$. 
The complete procedure for this sequential iterative method is as follows: 

\textbf{Step 1} Set initial guess for equilibrium energy density $e^{{\rm{0}}}$.

\textbf{Step 2} Solve the discretized form of the BTE (Eq. (\ref{eq:disretized_BTE})) subject to the boundary conditions to obtain the energy density ${e}$.
 
\textbf{Step 3} Calculate $e^{{\rm{0}}}$ based on Eq. (\ref{eq:BTE_get_e0}). 

\textbf{Step 4} Repeat Step 2 to Step 3 until convergence. 

The iteration stops when $\varepsilon_T  = {\sqrt {\sum\nolimits_i^{{N_{i}}} {{{\left( {T_{i}^n - T_{i}^{n + 1}} \right)}^2}/{N_{i}}} } }/{{{T_{\max }}}}<$ResidualTemp and $\varepsilon_q  = {\sqrt {\sum\nolimits_i^{{N_{i}}} {{{\left( {\left| {\bf{q}} \right|_i^n - \left| {\bf{q}} \right|_i^{n + 1}} \right)}^2}/{N_{i}}} } }/{{{{\left| {\bf{q}} \right|}_{\max }}}} <$ResidualFlux are both satisfied, where ${N_{i}}$ is the number of spatial cells. ResidualTemp and ResidualFlux are two input parameters in GiftBTE, which will be introduced in Section \ref{sec:workflow}. The temperature $T$ and heat flux $\bf{q}$ are obtained according to 
\begin{equation}
\begin{array}{c}
T = \frac{{\sum\limits_\alpha  {\sum\limits_\lambda  {{w_\alpha }{e_{\alpha ,\lambda }}} } }}{{\sum\limits_\lambda  {{C_\lambda }} }},\\
{\bf{q}} = \sum\limits_\alpha  {\sum\limits_\lambda  {{w_\alpha }{{\bf{v}}_\lambda }{e_{\alpha ,\lambda }}} } .
\end{array}
\label{eq:getTQ}
\end{equation}

In certain instances, such as cases involving near-diffusive conditions with a heat source, the convergence rate of the sequential iterative method can be remarkably sluggish. Consequently, GiftBTE offers a synthetic iterative scheme as a means to surmount this sluggish convergence rate.
Different from the sequential iterative method which updates the equilibrium energy density according to the equation (\ref{eq:BTE_get_e0}), the synthetic iterative scheme updates the equilibrium energy density according to a diffusion-type equation \cite{zhang2021fast}.
Following the derivation of Zhang et al., a diffusion-type equation can be obtained from the phonon BTE \citep{zhang2021fast}:

\begin{equation}
\begin{array}{*{20}{c}}
{e_\lambda ^{\rm{0}} = \frac{1}{{4\pi }}{C_\lambda }{T_L},}\\
{{k_{{\rm{bulk}}}}{\nabla ^2}{T_L} = \nabla  \cdot \left( {{{\bf{q}}_{{\rm{non - Fourier}}}}} \right) - \sum\limits_\lambda  {{{\dot Q}_\lambda }} ,}\\
{{{\bf{q}}_{{\rm{non - Fourier}}}} =  - \sum\limits_\lambda  {\sum\limits_\alpha  {({\tau _\lambda }v_\lambda ^2{{\bf{s}}_\alpha }{{\bf{s}}_\alpha } - {k_{{\rm{bulk}}}}\left( {{\tau _\lambda }\sum\limits_\lambda  {{C_\lambda }/{\tau _\lambda }} } \right){\bf{I}}) \cdot \nabla {e_{\alpha ,\lambda }}} } ,}
\end{array}
\end{equation}
where ${k_{\rm bulk}}$ is the bulk thermal conductivity expressed as ${k_{\rm bulk}}=\frac{1}{3}\sum\limits_{\lambda} {C_{\lambda}}v_{\lambda}^2{\tau _{\lambda}}$. This diffusion-type equation strengthens the coupling of phonon bands and therefore facilitates fast convergence from the diffusive regime to the ballistic regime \citep{zhang2021fast}. This equation is solved with the finite volume method:

\begin{equation}
\begin{array}{*{20}{c}}
{e_{i,\lambda }^{n + 1,{\rm{0}}} = \frac{1}{{4\pi }}{C_\lambda }T_{L,i}^{n + 1},}\\
{k_{{\rm{bulk}},i}}\sum\limits_{j \in N\left( i \right)} {{S_{ij}}{{\bf{n}}_{ij}} \cdot \nabla T_{L,ij}^{n + 1} = } \sum\limits_{j \in N\left( i \right)} {{S_{ij}}{{\bf{n}}_{ij}} \cdot \left( {{{\bf{q}}_{{\rm{non - Fourier}}}}} \right)_{ij}^{n + 1} - \sum\limits_\lambda  {{{\dot Q}_{i,\lambda }}} } ,
\end{array}
\label{eq:disretized_macro}
\end{equation}

\begin{equation}
\begin{array}{c}
({{\bf{q}}_{{\rm{non - Fourier}}}})_{ij}^{n + 1} =  - \sum\limits_\lambda  {\sum\limits_\alpha  {({\tau _{i,\lambda }}v_{i,\lambda }^2{{\bf{s}}_\alpha }{{\bf{s}}_\alpha } - {k_{{\rm{bulk}},i}}({\tau _{i,\lambda }}\sum\limits_\lambda  {{C_{i,\lambda }}/{\tau _{i,\lambda }}} ){\bf{I}}) \cdot \nabla e_{ij,\alpha ,\lambda }^{n + 1}} } .
\end{array}
\label{eq:disretized_non_fourier}
\end{equation}
To solve this equation, gradients on the faces of the cells are required. Non-orthogonal correction is adopted to obtain the gradient for unstructured meshes \citep{jaun1999numerical}.

The complete procedure of this method is as follows: 

\textbf{Step 1} Set initial guess for equilibrium energy density $e_{\lambda}^{{\rm{0}}}$.

\textbf{Step 2} Solve the discretized form of the BTE (Eq. (\ref{eq:disretized_BTE})) subject the boundary conditions to obtain the energy density ${e_{\lambda}}$.
 
\textbf{Step 3} Calculate ${{\bf{q}}_{{\rm{non - Fourier}}}}$ based on Eq. (\ref{eq:disretized_non_fourier}).

\textbf{Step 4} Solve the Eq. (\ref{eq:disretized_macro}) to update the $e_{\lambda}^{{\rm{0}}}$. 

\textbf{Step 5} Repeat Step 2 to Step 5 until convergence. 

The iteration stops when $\varepsilon  = {\sqrt {\sum\nolimits_i^{{N_{i}}} {{{\left( {T_{i}^n - T_{i}^{n + 1}} \right)}^2}/{N_{i}}} } }/{{{T_{\max }}}}<$ResidualTemp and $\varepsilon  = {\sqrt {\sum\nolimits_i^{{N_{i}}} {{{\left( {\left| {\bf{q}} \right|_i^n - \left| {\bf{q}} \right|_i^{n + 1}} \right)}^2}/{N_{i}}} } }/{{{{\left| {\bf{q}} \right|}_{\max }}}} <$ResidualFlux are both satisfied. The temperature $T$ and heat flux $\bf{q}$ are obtained according to Eq. (\ref{eq:getTQ}).

Considering that the synthetic iterative scheme necessitates the numerical solution of two equations, it may result in lower accuracy compared to the sequential iterative scheme. Hence, the sequential iterative scheme is recommended as the preferred iterative scheme. However, in cases where the sequential iterative method exhibits a considerably slow convergence rate, the synthetic iterative scheme can be employed as an alternative. GiftBTE also provides the choice of iterative scheme as an input parameter, which will be discussed in Sec. \ref{sec:workflow}.

\subsection{Implementation}

In addition to selecting numerical methods with high efficiency, the detailed implementation of these methods is crucial, including the selection of initial guesses, the choice of matrix solver, parallelization techniques, and algorithm organization. This subsection introduces the detailed implementation of the steady-state solver in GiftBTE.

To ensure faster convergence in the diffusive regime, the solution of the heat diffusion equation is used as the initial guess for the BTE solver \citep{romano2021openbte}. The heat diffusion equation is solved across the entire computational domain.
The steady-state heat diffusion equation is expressed as:
\begin{equation}
    \begin{array}{c}
{k_{\rm bulk}}{\nabla ^2}T_L =  -\sum\limits_{\lambda} {{{\dot Q}_{\lambda}}} ,\\
\end{array}
\end{equation}
This equation is also solved with the finite volume method: 
\begin{equation}
    {k_{{\rm bulk},i}}\sum\limits_{j \in N\left( i \right)} {{S_{ij}}{{\bf{n}}_{ij}} \cdot \nabla T_{L,ij}^{n + 1} = - \sum\limits_{\lambda} {{{\dot Q}_{i,\lambda}}} }. \label{eq:diff}
\end{equation}
To solve this equation, the gradient on the faces of the cells is required. The non-orthogonal correction is adopted to obtain the gradient for unstructured meshes \citep{jaun1999numerical}. 

The linear system resulting from the discretization of the heat diffusion equation can be formulated as follows:

\begin{equation}
{{\bf{A}}_{{\rm{diff}}}}{{\bf{T}}_{{L}}} = {{\bf{b}}_{{\rm{diff}}}},
 \end{equation}
 ${{\bf{A}}_{{\rm{diff}}}}$  is a sparse matrix of size $N_i\times N_i$, where $N_i$ is the number of spatial cells. The elements for this matrix is expressed as: 
\begin{equation}
\begin{array}{c}
{A_{ii,{\rm{diff}}}} =  - \sum\limits_{j \in N\left( i \right)} {{k_{{\rm{bulk}},i}}{S_{ij}}{{\bf{n}}_{ij}} \cdot \frac{{{{\bf{n}}_{ij}}}}{{{{\bf{n}}_{ij}} \cdot {{\bf{d}}_{ij}}}},} \\
{A_{ij,{\rm{diff}}}} = {k_{{\rm{bulk}},i}}{S_{ij}}{{\bf{n}}_{ij}} \cdot \frac{{{{\bf{n}}_{ij}}}}{{{{\bf{n}}_{ij}} \cdot {{\bf{d}}_{ij}}}},
\end{array}
\end{equation}
where ${\bf{d}}_{ij}$ is the vector that connects the two cells. ${\bf{b}}$ contains contributions from the right-hand side of equation (\ref{eq:diff}) and the non-orthogonal correction:
  \begin{equation}
  \begin{aligned}
{b_{i,{\rm{diff}}}} = \sum\limits_{j \in N\left( i \right)} {{k_{{\rm{bulk}},i}}{S_{ij}}\nabla {T_{L,f,ij}}\left( {{{\bf{n}}_{ij}} - {{\bf{n}}_{ij}} \cdot \frac{{{{\bf{n}}_{ij}}}}{{{{\bf{n}}_{ij}} \cdot {{\bf{d}}_{ij}}}}} \right)}  - \sum\limits_\lambda  {{{\dot Q}_{i,\lambda }}} .
  \end{aligned}
 \label{eq:B_HF}
  \end{equation}

$\nabla {T_{L,f,ij}}$ denotes the gradient of lattice temperature along the face, which is computed from the lattice temperature values at the vertices. The lattice temperature values at vertices are determined by interpolating the cell-centered values. After obtaining the temperature field through the heat diffusion equation, the equilibrium energy density can be determined using equation (\ref{eq:BTE_get_e0}). Subsequently, the phonon BTE is solved.

The iterative linear system obtained from the discretized form of the phonon BTE (Eq. (\ref{eq:disretized_BTE})) can be expressed as follows:

\begin{equation}
{{\bf{A}}_{\alpha,\lambda}}\Delta {\bf{e}}_{\alpha,\lambda}^{n + 1} = {\bf{b}}_{\alpha,\lambda  }^n.
\label{Linear system}
\end{equation}
Here, the stiffness matrix for each band and direction is denoted by ${{\bf{A}}_{\alpha,\lambda}}$, which is a sparse matrix with the size of $N_i\times N_i$. At row $i$, the matrix elements are given by:
\begin{equation}
\begin{array}{c}
{A_{ii,\alpha ,\lambda }} = \frac{1}{{{\tau _{i,\lambda }}}} + \sum\limits_{j \in N\left( i \right)} {\frac{1}{{{V_i}}}{{\bf{v}}_{i,\lambda }} \cdot {{\bf{n}}_{ij}}{\rm{\Theta }}({{\bf{v}}_{i,\lambda }} \cdot {{\bf{n}}_{ij}}){S_{ij}}} ,\\
{A_{ij,\alpha ,\lambda }} = \frac{1}{{{V_i}}}{{\bf{v}}_{i,\lambda }} \cdot {{\bf{n}}_{ij}}{\rm{\Theta }}( - {{\bf{v}}_{i,\lambda }} \cdot {{\bf{n}}_{ij}}){S_{ij}}.
\end{array}
\end{equation}
$\Theta$ is the Heaviside function.
The right-hand-side of Eq. (\ref{eq:disretized_BTE}) is denoted by ${\bf{b}}^n$, which is expressed as:
\begin{equation}
\begin{array}{c}
b_{i,\alpha,\lambda}^n =  - \sum\limits_{j \in N\left( i \right)} {\frac{1}{{{V_i}}}{{\bf{v}}_{i,\lambda }} \cdot {{\bf{n}}_{ij}}\Theta ({{\bf{v}}_{i,\lambda }} \cdot {{\bf{n}}_{ij}}){S_{ij}}\left( {e_{i,\alpha ,\lambda }^n + \nabla e_{i,\alpha ,\lambda }^n \cdot {{\bf{l}}_{i,ij}}} \right)} \\
 - \sum\limits_{j \in N\left( i \right)} {\frac{1}{{{V_i}}}{{\bf{v}}_{i,\lambda }} \cdot {{\bf{n}}_{ij}}\Theta ( - {{\bf{v}}_{i,\lambda }} \cdot {{\bf{n}}_{ij}}){S_{ij}}\left( {e_{j,\alpha ,\lambda }^n + \nabla e_{j,\alpha ,\lambda }^n \cdot {{\bf{l}}_{i,ij}}} \right)}  - \frac{{e_{i,\alpha ,\lambda }^n}}{{{\tau _{i,\lambda }}}} + \frac{{e_{i,\lambda }^{n,{\rm{0}}}}}{{{\tau _{i,\lambda }}}} + {{\dot Q}_{i,\lambda }}.
\end{array}
\label{eq:B_BTE}
 \end{equation}

In these expressions, the gradient in each cell needs to be calculated from the energy density. The least squares method is adopted to obtain the gradient. 
\begin{equation}
\nabla {e_i} = {{\bf{X}}^{ - 1}}{\bf{y}},
 \end{equation}
where $\bf{X}$ is related to the coordinates of cell $i$ and its neighbors $j \in N\left( i \right)$. For each neighbor $j$, $y_j$ is defined as $e_j-e_i$.

For the synthetic iterative scheme, to obtain the equilibrium energy density ${e_{i,\lambda}^{n,{\rm{0}}}}$, equation (\ref{eq:disretized_macro}) is solved, resulting in another linear system:
\begin{equation}
{{\bf{A}}_{{\rm{syn}}}}{{\bf{T}}_L} = {{\bf{b}}_{{\rm{syn}}}},
 \end{equation}
where ${\bf{A}}_{\rm syn}$ is the stiffness matrix for the diffusion-type equation (\ref{eq:disretized_macro}). It is a sparse matrix of size $N_i\times N_i$. The matrix elements for row $i$ are given by:
 \begin{equation}
 \begin{array}{c}
{A_{ii{\rm{,syn}}}} =  - \sum\limits_{j \in N\left( i \right)} {{k_{{\rm{bulk}},i}}{S_{ij}}{{\bf{n}}_{ij}} \cdot \frac{{{{\bf{n}}_{ij}}}}{{{{\bf{n}}_{ij}} \cdot {{\bf{d}}_{ij}}}}} ,\\
{A_{ij{\rm{,syn}}}} = {k_{{\rm{bulk}},i}}{S_{ij}}{{\bf{n}}_{ij}} \cdot \frac{{{{\bf{n}}_{ij}}}}{{{{\bf{n}}_{ij}} \cdot {{\bf{d}}_{ij}}}},
\end{array}
 \end{equation}
 ${\bf{b}}$ contains contributions from the right-hand side of equation (\ref{eq:disretized_macro}) and the non-orthogonal correction:
  \begin{equation}
\begin{array}{c}
{b_i} = \sum\limits_{j \in N\left( i \right)} {{k_{{\rm{bulk}},i}}{S_{ij}}\nabla {T_{L,f,ij}}\left( {{{\bf{n}}_{ij}} - {{\bf{n}}_{ij}} \cdot \frac{{{{\bf{n}}_{ij}}}}{{{{\bf{n}}_{ij}} \cdot {{\bf{d}}_{ij}}}}} \right)} \\
 + \sum\limits_{j \in N\left( i \right)} {{S_{ij}}{{\bf{n}}_{ij}} \cdot \left( {{{\bf{q}}_{{\rm{non - Fourier}}}}} \right)_{ij}^{n + 1} - \sum\limits_\lambda  {{{\dot Q}_{i,\lambda }}} } .
\end{array} 
 \label{eq:B_HF}
  \end{equation}
${{{\bf{q}}_{{\rm{non - Fourier}}}}}$ is calculated according to equation (\ref{eq:disretized_non_fourier}). The gradient of energy density on each face is determined using the non-orthogonal correction, expressed as:
\begin{equation}
 \nabla {e_{ij}} = \frac{{{e_i} - {e_{ij}}}}{{{{\bf{d}}_{ij}}}} + \nabla {e_{f,}}_{ij}.
\end{equation}
Here, $\nabla {e_{f,ij}}$ represents the gradient along the face, which is computed from the energy density values at the vertices. The energy density values at vertices are determined by interpolating the cell-centered values.

\begin{algorithm} [!htb]
\caption{Overall implementation of the steady-state solver}\label{alg:cap}
\begin{algorithmic}[1]
\State Inputs: Phonon properties, meshes, et al.
\State Sample phonon bands $\lambda$ and directions $\bf{s}_{\alpha}$ according to Eq. (\ref{eq:banddis}) and Eq. (\ref{eq:directiondis})
\State $\bf{T}_{\it{L}} \gets \bf{A_{\rm diff}}{{\bf{T}_{\it{L}}}} = {\bf{b}_{\rm diff}}$. \Comment{LU solver}
\State Compute ${\bf{A}}_{\alpha,\lambda}$ for all bands and directions, ${\bf{A}}_{\rm syn}$, and  $\bf{X}$ for all cells.
\While {$\varepsilon_T >$ ResidualTemp or $\varepsilon_q>$ ResidualFlux}.
\For {all $\lambda$ and $\bf{s}_{\alpha}$ }
\For {$i = 1 : {N_{i}} $}

\State $\nabla {e_i} = {{\bf{X}}^{ - 1}}{\bf{y}}$.

\EndFor

\State  ${\bf{b}}_{\alpha,\lambda}^n \gets$ Equation (\ref{eq:B_BTE}).

\State $\Delta {\bf{e}}_{\alpha ,\lambda }^{n + 1} \leftarrow {{\bf{A}}_{\alpha ,\lambda }}\Delta {\bf{e}}_{\alpha ,\lambda }^{n + 1} = {\bf{b}}_{\alpha ,\lambda }^n.$ \Comment{LU solver}
\State ${\bf{e}}_{\alpha ,\lambda }^{n + 1} = \Delta {\bf{e}}_{\alpha ,\lambda }^{n + 1} + {\bf{e}}_{\alpha ,\lambda }^n.$
\State $T_L^{n + 1} \leftarrow {T_L} = \frac{{\frac{1}{{4\pi }}\sum\limits_\alpha  {\sum\limits_\lambda  {\frac{{{w_\alpha }{e_{\alpha ,\lambda }}}}{{{\tau _\lambda }}}} } }}{{\sum\limits_\lambda  {\frac{{{C_\lambda }}}{{{\tau _\lambda }}}} }}.$
\For {$i = 1 : {N_{i}} $}
\For{$j \in N(i)$}
\State $({{\bf{q}}_{{{\rm non - Fourier}}}})_{ij}^{n+1} \gets$ Eq.   (\ref{eq:disretized_non_fourier}). \Comment{Only for synthetic iterative scheme}
\EndFor
\EndFor
\EndFor

\State ${\bf{b}}_{\rm syn} \gets$ Equation (\ref{eq:B_HF}) \Comment{Only for synthetic iterative scheme}
\State $\bf{T}_{\it{L}} \gets {\bf{A}}_{\rm syn}{{\bf{T}}_{\it{L}}} = {\bf{b}}_{\rm syn}.$ \Comment{Only for synthetic iterative scheme}
\State 
${\bf{e}}_\lambda ^{n + {\rm{1}},{\rm{0}}} = \frac{1}{{4\pi }}{{\bf{C}}_\lambda }{{\bf{T}}_L}.$
 
\State $\varepsilon_T  = \frac{{\sqrt {\sum\nolimits_i^{{N_{i}}} {{{\left( {T_{i}^n - T_{i}^{n + 1}} \right)}^2}/{N_{i}}} } }}{{{T_{\max }}}}.$
\State $\varepsilon_q  = \frac{{\sqrt {\sum\nolimits_i^{{N_{i}}} {{{\left( {\left| {\bf{q}} \right|_i^n - \left| {\bf{q}} \right|_i^{n + 1}} \right)}^2}/{N_{i}}} } }}{{{{\left| {\bf{q}} \right|}_{\max }}}}.$
\EndWhile
\State $T = \frac{{\sum\limits_\alpha  {\sum\limits_\lambda  {{w_\alpha }{e_{\alpha ,\lambda }}} } }}{{\sum\limits_\lambda  {{C_\lambda }} }}.$
\State ${\bf{q}} = \sum\limits_\alpha  {\sum\limits_\lambda  {{w_\alpha }{{\bf{v}}_\lambda }{e_{\alpha ,\lambda }}} } .$
\end{algorithmic}
\label{alor}
\end{algorithm}

In the above expression, the stiffness matrix for each band and each direction ${{\bf{A}}_{\alpha ,\lambda }}$, the stiffness matrix for the diffusion-type equation ${\bf{A}_{\rm syn}}$, and the matrix $\bf{X}$ used for the least squares computation, remain unchanged throughout the iterations. These matrices can then be computed in advance, prior to the start of the iteration. In addition to computing the matrices themselves, the inversion of the matrix $\bf{X}$ and the preconditioning of the matrix can also be performed beforehand, before the iterations commence. GiftBTE computes the LU factorization of the stiffness matrix only once and utilizes it for different residuals during each iteration. The LU factorization is achieved using the Eigen package \citep{eigenweb}. The column approximate minimum degree ordering is employed to reduce memory consumption \citep{eigenweb}. The overall implementation is summarized in Algorithm \ref{alg:cap}.

In the current solver, MPI parallelization is employed. The loops corresponding to steps from Line 4 to Line 16 iterate over the directions, bands, and cells. Prior research has demonstrated the effectiveness of a particular parallelization strategy \citep{hu2022ultra}: when the number of CPU cores is smaller than the number of directions, the direction loops are expanded, allowing different CPUs to simultaneously compute different directions. Conversely, when the number of CPU cores exceeds the number of directions, the band loops are further expanded. This parallelization strategy is also implemented in GiftBTE. The steps from Line 16 to Line 20 exclusively involve cells. Hence, the cell loops are expanded for these specific steps.

\begin{figure} [!htb]
    	\centering
    	\includegraphics[width=1\textwidth]{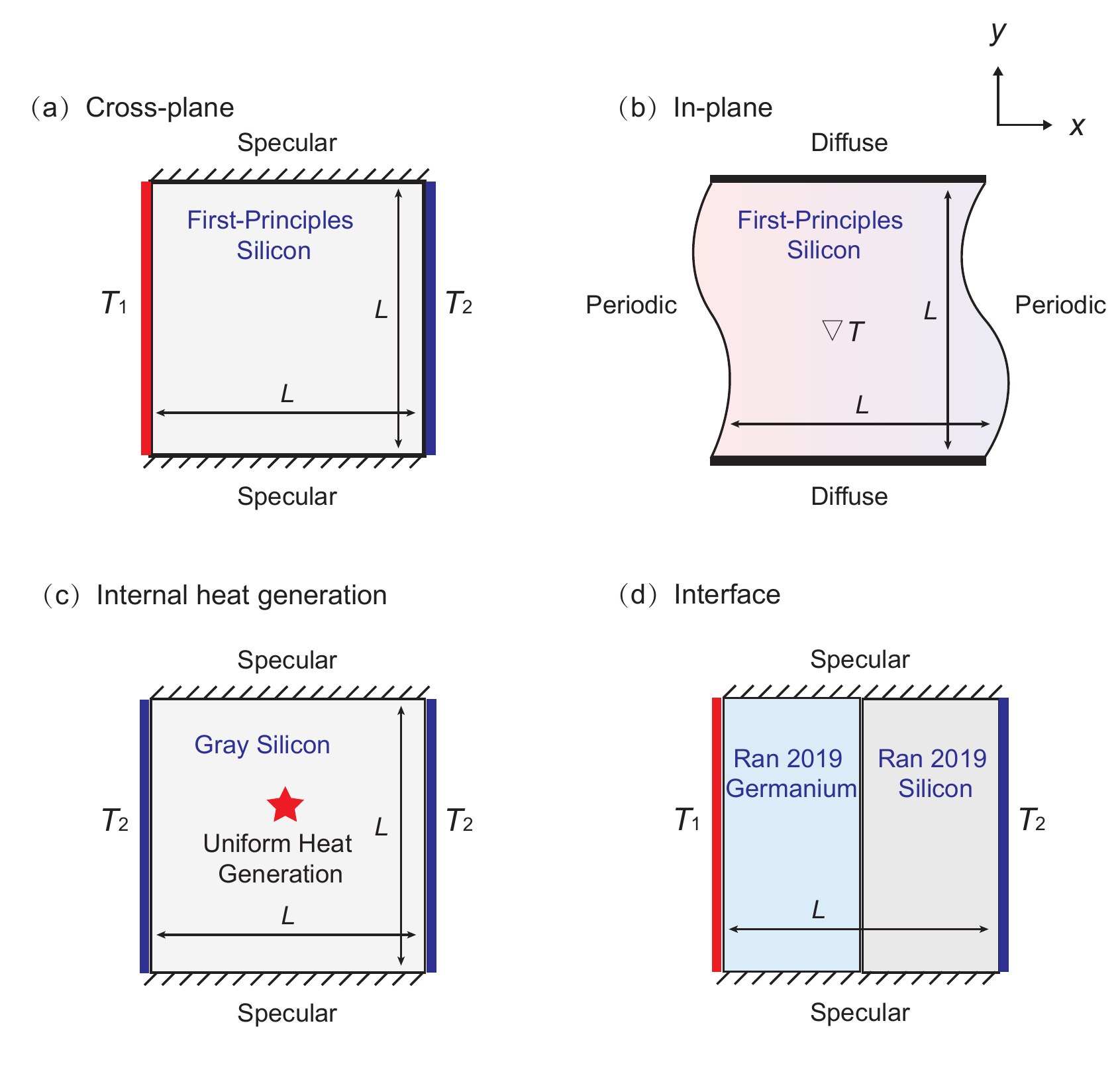}
    	\caption{
    	The toy problems to verify the accuracy of our numerical implementation. (a) cross-plane heat conduction; (b) in-plane heat conduction; (c) internal heat generation; and (d) heat conduction across an Ge-Si interface.
    	}
    	\label{fig:verifycase}
\end{figure}

\subsection{Verification}

Some toy problems that have analytical solutions or numerical results from previous references are computed to verify the correctness of the steady-state solver in the GiftBTE. Four problems including cross-plane heat conduction in Si thin films, in-plane heat conduction in Si thin films, internal heat generation in Si thin films, and thermal transport across the Ge-Si interface have been selected (as depicted in Fig. \ref{fig:verifycase}). The analytical solutions \citep{hua2015semi,chen2005nanoscale} are used to verify the former two problems, while numerical results in references \citep{hua2015semi} and \citep{ran2019efficiency} are used to verify the latter two. All elements in the solver, including all boundary conditions and the volumetric heat generation term, are incorporated in these four problems. A square simulation cell with a length of $L$ is used for all four problems. In cross-plane heat conduction (Fig. \ref{fig:verifycase}(a)), thermalizing boundary conditions with temperatures $T_{\rm 1}$ and $T_{\rm 2}$ are set at the left and right boundaries, respectively. The top and bottom boundaries are set as specularly reflecting boundary conditions. In in-plane heat conduction (Fig. \ref{fig:verifycase}(b)), periodic boundary conditions are set at the left and right boundaries with a temperature difference, which induces a temperature gradient over the x direction. The top and bottom boundaries are set as diffusely reflecting boundary conditions. For these two problems, first-principles Si is used as a material. The phonon properties of Si are obtained through first-principles calculations, employing the anharmonic lattice dynamics method implemented in the ALAMODE package \citep{tadano2014anharmonic}. Our first-principles calculations employ a 50×50×50 grid of q-points to sample the Brillouin zone, resulting in 50×50×50×6 phonon modes. The calculated bulk thermal conductivity of Si is found to be 145 W/m-K, which is in excellent agreement with literature values \citep{lindsay2018survey,tadano2014anharmonic}. The phonon properties including heat capacity, group velocity and relaxation time serve as inputs of GiftBTE. 8 phonon bands with 64 directions, which are the convergent phonon modes using the implemented band discretization and directional discretization schemes discussed in Sec. 3.1 and 3.2, are used. In internal heat generation (Fig. \ref{fig:verifycase}(c)), thermalizing boundary conditions with the same temperature $T_{\rm 2}$ are set at the left and right boundaries, while the top and bottom boundary are set as specularly reflecting boundary conditions. A uniform heat generation over the entire simulation cell is also set. Since the reference uses the gray model \citep{hua2015semi}, gray model Si (one band with the group velocity of 2677 m/s, the relaxation time of 39.9 ps, and the heat capacity of 1627600 J/(Km$^{3}$)) is used as the material. For the thermal transport across Si-Ge interface (Fig. \ref{fig:verifycase}(d)), the thermalizing boundary conditions with temperatures $T_{\rm 1}$ and $T_{\rm 2}$ are set at the left and right boundaries, respectively. The top and bottom boundaries are set as specularly reflecting boundary conditions. The left-side material is set as Ge, and the right-side material is set as Si. The phonon properties are the same as that in the Ref. \citep{ran2019efficiency}.

\begin{figure}
    	\centering
    	\includegraphics[width=1\textwidth]{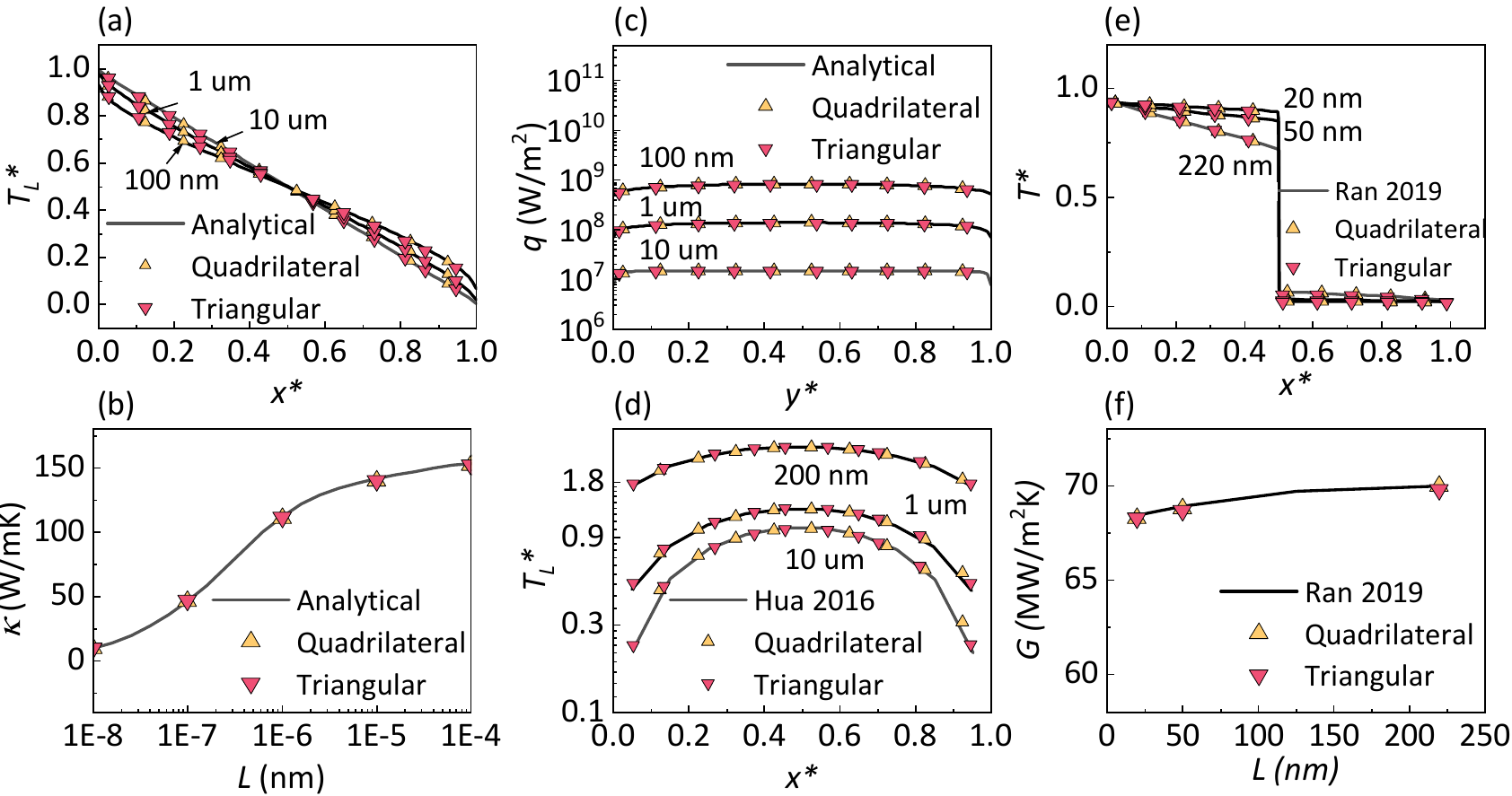}
    	\caption{
    	The numerical results of four toy problems verified with analytical solutions \cite{chen2005nanoscale,hua2015semi} or the numerical results in references \cite{ran2019efficiency,hua2016transient}.)
    	}
    	\label{fig:verifyresult}
\end{figure}

For all of four problems, both quadrilateral and triangular meshes are tested, as the present solver is an unstructured solver. Upon completing convergence tests, 400 quadrilateral meshes or 2900 triangular meshes are utilized to discretize the computational domain. The results obtained by GiftBTE (adopting synthetic iterative scheme) are depicted in Fig. \ref{fig:verifyresult}. In Fig. \ref{fig:verifyresult}, $x^*$ and $y^*$ represent the dimensionless coordinates which are defined as $x/L$ and $y/L$. The Fig. \ref{fig:verifyresult}(a) shows the distribution of dimensionless lattice temperature $T_L^*$ in the cross-plane problem. The dimensionless lattice temperature $T_L^*$ is defined as ${T_L - {T_2}}/{{{T_1} - {T_2}}}$. The Fig. \ref{fig:verifyresult}(b) shows the effective thermal conductivity $\kappa$ of the cross-plane problem, which is defined as ${{q_x}}{{({T_1} - {T_2})/L}}$. The Fig. \ref{fig:verifyresult}(c) shows the heat flux distribution of the in-plane problem. The Fig. \ref{fig:verifyresult}(d) shows the dimensionless lattice temperature $T_L^*$ of the internal heat generation problem, which is defined as $8{k_{{\rm{bulk}}}}(T_L - {T_2})/(4\pi \dot q L_x^2)$. Fig.  \ref{fig:verifyresult}(e) shows the dimensionless temperature distribution $T^*$, which is defined as $(T - {T_2})/({T_1} - {T_2})$. The Fig. \ref{fig:verifyresult}(f) shows the interfacial thermal conductance $G$, which is defined as $G/{{{T_{{\rm{left}}}} - {T_{{\rm{right}}}}}}$, where the  ${T_{{\rm{left}}}}$ and ${T_{{\rm{right}}}}$ is the temperature in the left side of the interface and the right side of the interface, respectively. For all four problems, the numerical results are consistent with the analytical solutions or the numerical results found in the reference (Fig. \ref{fig:verifyresult}). The results obtained with employing the sequential iterative scheme also demonstrate good agreement with the analytical solutions or numerical results, but are not presented here for the sake of simplicity. All these cases have been listed as examples of the GiftBTE packages.

\section{Transient solver} \label{sec:transient} 

The transient phonon BTE is expressed as:

\begin{equation}
 \frac{{\partial e}}{{\partial t}} + {\bf{v}} \cdot \nabla e =  - \frac{{e - {e^0}}}{\tau } + \dot Q.
\end{equation}

In GiftBTE, this equation is solved by the explicit DOM. To numerically solve this equation using DOM, one needs band discretization, directional discretization, time discretization and spatial discretization to transform the integro-differential equation into algebraic equations and then solve all algebraic equations. For the transient solver, GiftBTE adopts the same band and directional discretization as that used in the steady-state solver (Sec. \ref{sec:steady}). For time and spatial discretization, GiftBTE adopts the forward Euler scheme (explicit scheme) and finite-volume method with second-order accuracy.

\subsection{Time and spatial discretization}
By adopting band discretization and directional discretization in Sec. 3.1 and 3.2, the transient phonon BTE becomes a partial differential equation:

\begin{equation}
\frac{{\partial {e_{\alpha ,\lambda }}}}{{\partial t}} + {{\bf{v}}_\lambda } \cdot \nabla {e_{\alpha ,\lambda }} =  - \frac{{{e_{\alpha ,\lambda }} - e_\lambda ^{\rm{0}}}}{{{\tau _\lambda }}} + {\dot Q_\lambda }.
\end{equation}

By using the forward Euler scheme and finite-volume method to discrete this equation, we get:
\begin{equation}
\begin{aligned}
 e_{i,\alpha ,\lambda }^{n + 1} = e_{i,\alpha ,\lambda }^n - \Delta t\sum\limits_{j \in N\left( i \right)} {e_{ij,\alpha ,\lambda }^n} {{\bf{v}}_{i,\lambda }} \cdot {{\bf{n}}_{ij}}{S_{ij}} + {V_i}\Delta t( - \frac{{e_{i,\alpha ,\lambda }^n - e_{i,\lambda }^{n,0}}}{{{\tau _{i,\lambda }}}} + {\dot Q_{i,\lambda }})
 \end{aligned}
 \label{eq:disretized_transientBTE}
\end{equation}
where $\Delta{t}$ is the time step. $n$ denotes the index of time $t_n=n\Delta{t}$. The energy density ${e_{ij,\bf{s}}}$ on the face is calculated from the cell-centered energy density and the gradient of the energy density by
\begin{equation}
{e_{ij}}={e_m} + \nabla {e_m} \cdot {{\bf{l}}_{m,ij}},
\end{equation}
where cell $m$ is the cell in the upwind direction, and ${{\bf{l}}_{m,ij}}$ is the vector from the center of cell $m$ to the center of the face $ij$. The gradient $\nabla {e_i}$ is calculated by the least squares method. 
\par
The complete procedure for this method is as follows: 

\textbf{Step 1} Set equilibrium energy density $e^{{\rm{0}}}$ and energy density $e$ at $t_{n}$.

\textbf{Step 2} Solve the discretized form of the BTE (Eq. (\ref{eq:disretized_transientBTE})) subject the boundary conditions to obtain the energy density ${e^{n+1}}$.
 
\textbf{Step 3} Calculate equilibrium energy density ${e^{0,n+1}}$ based on Eq. (\ref{eq:disretized_non_fourier}).

\textbf{Step 4} Repeat Step 2 to Step 3 until reach the desired simulation time or reach steady-state. 

As a explicit method, the time step $\Delta{t}$ is limited by the Courant-Friedrichs-Lewy (CFL) condition
\begin{equation}
 \Delta{t}=\beta \frac{\Delta x_{\rm min}}{v}. 
\end{equation}
where $\Delta x_{\rm min}$ is the minimum cell size and $0<\beta<1$ is the CFL number.

The overall implementation is summarized in Algorithm \ref{alg:tra}. The time step $\Delta t$, the desired simulation time  (TotalTime), initial equilibrium energy density $e^{{\rm{0}}}$ and energy density $e$ also serve as inputs of GiftBTE, which will be discussed in Sec. \ref{sec:workflow}.

\begin{algorithm} [!htb]
\caption{Overall implementation of the transient solver}\label{alg:tra}
\begin{algorithmic}[1]
\State Inputs: Phonon properties, meshes, et al.
\State Sample phonon bands $\lambda$ and directions $\bf{s}_{\theta,\varphi}$ according to Eq. (\ref{eq:banddis}) and Eq. (\ref{eq:directiondis})
\State Initial $e$ and $e_{0}$ for all bands, directions,and cells at time step $n=0$.
\While {$\varepsilon_T >$ ResidualTemp or $\varepsilon_q>$ ResidualFlux and $n{\Delta}t<$TotalTime}.
\For {all $\lambda$ and $\bf{s}_{\theta,\varphi}$ }
\For {$i = 1 : {N_{cell}} $}

\State $\nabla {e_i} = {{\bf{X}}^{ - 1}}{\bf{y}}$.

\EndFor
\For {$i = 1 : {N_{i}} $}

\State $e_{i,\alpha,\lambda}^{n+1}\gets$ Equation (\ref{eq:disretized_transientBTE})

\EndFor
\State $T_L^{n + 1} \leftarrow {T_L} = \frac{{\frac{1}{{4\pi }}\sum\limits_\alpha  {\sum\limits_\lambda  {\frac{{{w_\alpha }{e_{\alpha ,\lambda }}}}{{{\tau _\lambda }}}} } }}{{\sum\limits_\lambda  {\frac{{{C_\lambda }}}{{{\tau _\lambda }}}} }}.$

\EndFor
\State 
${\bf{e}}_\lambda ^{n + {\rm{1}},{\rm{0}}} = \frac{1}{{4\pi }}{{\bf{C}}_\lambda }{{\bf{T}}_L}$
\State $\varepsilon_T  = \frac{{\sqrt {\sum\nolimits_i^{{N_{i}}} {{{\left( {T_{i}^n - T_{i}^{n + 1}} \right)}^2}/{N_{i}}} } }}{{{T_{\max }}}}.$
\State $\varepsilon_q  = \frac{{\sqrt {\sum\nolimits_i^{{N_{i}}} {{{\left( {\left| {\bf{q}} \right|_i^n - \left| {\bf{q}} \right|_i^{n + 1}} \right)}^2}/{N_{i}}} } }}{{{{\left| {\bf{q}} \right|}_{\max }}}}.$
\EndWhile
\State $T = \frac{{\sum\limits_\alpha  {\sum\limits_\lambda  {{w_\alpha }{e_{\alpha ,\lambda }}} } }}{{\sum\limits_\lambda  {{C_\lambda }} }}.$
\State ${\bf{q}} = \sum\limits_\alpha  {\sum\limits_\lambda  {{w_\alpha }{{\bf{v}}_\lambda }{e_{\alpha ,\lambda }}} } .$
\end{algorithmic}
\label{alor}
\end{algorithm}

\subsection{Verification}

\begin{figure}[!htb]
    	\centering
    	\includegraphics[width=0.8\textwidth]{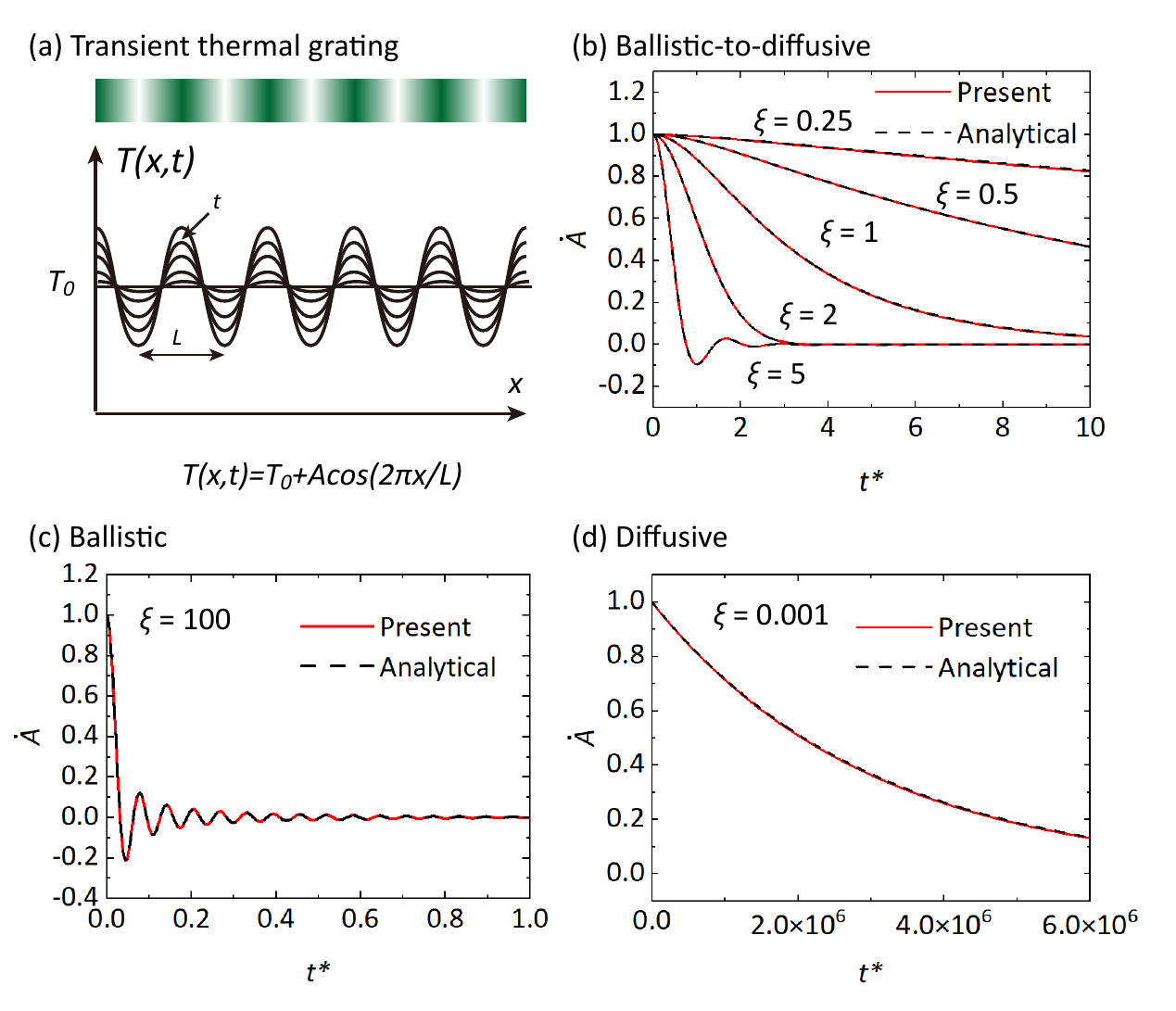}
    	\caption{The numerical results of one-dimensional transient thermal grating verified with analytical solutions (a) Schematic of one-dimensional transient thermal grating. (b) Decaying of the dimensionless amplitude of temperature variation $\dot{A}=A/A_0$ against dimensionless time $t^*=t/\tau$ at ballistic-to-diffusive regime ${\xi}=2{\pi}{\rm Kn}$. (c) Decaying of the dimensionless amplitude of temperature variation $\dot{A}=A/A_0$ against dimensionless time $t^*=t/\tau$ at ballistic regime.  (d)  Decaying of the dimensionless amplitude of temperature variation $\dot{A}=A/A_0$ against dimensionless time $t^*=t/\tau$ at diffusive regime.}
    	\label{fig:TTG}
\end{figure}

The transient solver is verified with the one-dimensional transient thermal grating (TTG) problem (as depicted in the Fig.\ref{fig:TTG}(a)), which has an analytical solution for the gray model (only consider one phonon band $\lambda$) 
 \citep{collins2013non}. In this problem, a spatially sinusoidal temperature variation is set initially,
\begin{equation}
    T(x,t) = T_0+A_{0}{\rm cos}(2{\pi}x/L) 
\end{equation}
where $T_0$ is the background temperature, $A_0$ is the amplitude of the temperature variation, and $L$ is the grating period. All phonons are in their equilibrium state at the temperature of $T$, i.e, the energy density $e$ for all bands and directions and their equilibrium energy density $e^0$ are $C_\lambda T/4\pi$. As time progresses, the amplitude of the spatially sinusoidal temperature variation decreases. We study one period and set the left and right boundaries as periodic boundary conditions. After conducting a convergence test, 1000 meshes and 512 directions are adopted. We examine the results from the ballistic regime to the diffusive regime. Fig.\ref{fig:TTG}(b-d) shows the amplitude of the temperature variation $\dot{A}=A/A_0$ at different times. $\xi$ is defined as ${\xi}=2{\pi}{Kn}$, where ${Kn}$ is the Knudsen number defined as ${Kn}=v{\tau}/L$. $t^*$ represents the dimensionless time, defined as $t/\tau$. 
It can be seen that the numerical solution for different Knudsen numbers agrees well with the analytical solution for all cases. All these cases have been listed as examples in the GiftBTE package.

\section{Workflow} \label{sec:workflow}

\begin{figure}[!htb]
    	\centering
    	\includegraphics[width=1\textwidth]{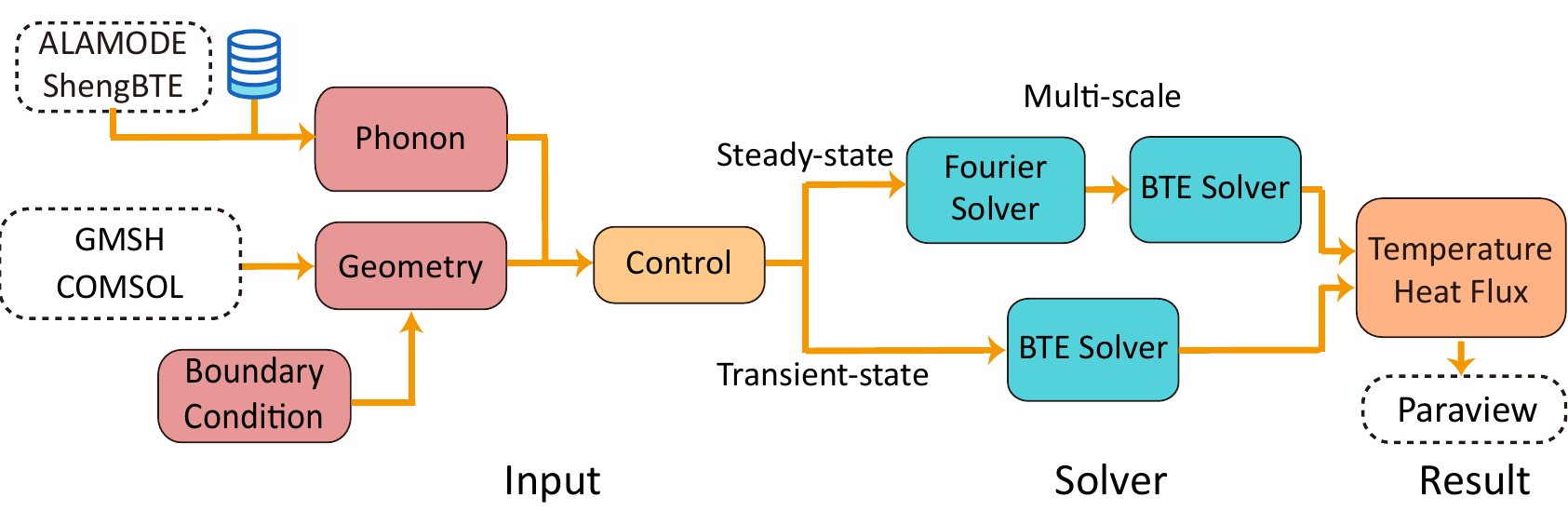}
    	\caption{Workflow of the GiftBTE.}.
    	\label{fig:method-figure1}
\end{figure}

The workflow sketch of GiftBTE is depicted in Fig. \ref{fig:method-figure1}. There are three main parts in GiftBTE: Input, Solver and Result. The Solver parts contain the steady-state solver (Sec. \ref{sec:steady}) and transient solver (Sec. \ref{sec:transient}). We have mentioned that some parameters serve as inputs for GiftBTE in Sec. \ref{sec:steady} and Sec. \ref{sec:transient}. These inputs are defined by three modules in the Input part: namely Control, Geometry, Phonon. The Input part for phonon properties provides the interface with first-principles calculation packages ShengBTE \cite{li2014shengbte} and ALAMODE \cite{tadano2014anharmonic}. The Input part for meshes provides the interface with two external mesh generators Gmsh \cite{geuzaine2009gmsh} and COMSOL. The Result part provides the interface with Paraview (www.paraview.org).
\subsection{Control}
In the Control module, users are required to choose between steady-state or transient solvers. For the steady-state solver, users are required to choose between the sequential iterative scheme and the synthetic iterative scheme. For steady-state, users also need to define the convergence criterion for temperature and heat flux and the maximum number of iterations. For the transient solver, users need to define the desired simulation time, time step and provide an initial temperature profile. The reference temperature is also defined in this module. These parameters are defined by the following commands in the input file named CONTROL: 
\begin{itemize}
    \item \textbf{State}:  This command defines which solver is adopted. "Transient" represents the transient solver, while "Steady" corresponds to the steady state solver.
    \item \textbf{IterativeScheme (for steady-state)}: This command specifies which scheme is employed. "Sequential" refers to adopting the sequential iterative scheme, and "synthetic" refers to the synthetic iterative scheme.
    \item \textbf{MaxNumIter (for steady-state)}: This command sets the maximum number of iterations. The default value is "10000".
    \item \textbf{ResidualTemp}: This command establishes the temperature criteria for convergence in the steady state solver and criteria for reaching steady-state in the transient solver. The suggested value is "1e-5."
    \item \textbf{ResidualFlux}: This command establishes the heat flux criteria for convergence in steady state solver or criteria for reaching steady-state in transient solver. The suggested value is "1e-3."
    \item \textbf{TotalTime (for transient)}: This command sets the desired simulation time of the transient solver.
    \item \textbf{TimeStep (for transient)}: This command sets the time step of the transient solver.
    \item \textbf{InitialTemperatureFile (for transient)}: This command specifies the path of the initial temperature file, which sets the initial temperature profile of the transient solver.
     \item \textbf{Tref}: This command sets the reference temperature.
\end{itemize}

\subsection{Geometry}
For both the steady-state solver and the transient solver, the meshes, boundary conditions and heat generation distributions serve as inputs.  The following commands are defined in the input file named GEOMETRY:

\begin{itemize}
\item \textbf{GeometryDimension}: This command defines the dimension of the computational domain. "1" denotes a one dimensional domain. "2" denotes a two dimensional domain. "3" denotes a three dimensional domain.
\item \textbf{ScaleX}: Scale in the $x$ direction. The unit is m.
\item \textbf{ScaleY}: Scale in the $y$ direction. The unit is m.
\item \textbf{ScaleZ}: Scale in the $z$ direction. The unit is m.

Note: the real size of the mesh adopted by the solver is the size of the input mesh multiplied by the scale.

\item \textbf{MeshType}: This command defines the type of mesh files. "MSH" refers to the .msh mesh file from GMSH \citep{geuzaine2009gmsh}, while "COMSOL" refers to the .mphtxt file from COMSOL.
\item \textbf{MeshFile}: This command specifies the path of the mesh file.
\item \textbf{HeatType}: This command specifies the type of heat generation file. Users can provide the total heat generation term $\sum\limits_{\lambda} {{{\dot Q}_{\lambda}}} $ in each spatial point by using the "COORDINATE" format defined by GiftBTE. More details about the format of the heat generation file can be found in the GiftBTE manual.
\item \textbf{HeatFile}: This command specifies the path of the heat source file.
\item \textbf{BCFile}: This command specifies the path of the boundary condition file. In the boundary condition file, the five types of boundary conditions discussed in Sec. \ref{sec:BC} can be defined. More details about the format of the boundary condition file can be found in the GiftBTE manual.
\end{itemize}

\subsection{Phonon}
For both the steady-state solver and the transient solver, the phonon properties serve as inputs. 
The Phonon module reads the input file named PHONON. In this file, several commands need to be defined.

\begin{itemize}
\item \textbf{MaterialNumber}: This command defines how many kinds of materials are involved in this computation.
\item \textbf{BandNumber}: This command defines bands $\lambda$ that are discretized in this computation.
\item \textbf{MaterialFile}: This command defines the type and path of the phonon properties file. "ALAMODE" is used to adopt the interface with ALAMODE \citep{tadano2014anharmonic}, which reads the phonon lifetime files from ALAMODE. "ShengBTE" is used to adopt the interface with ShengBTE \citep{li2014shengbte}, which reads BTE.omega, BTE.qpoints\_full, BTE.v, BTE.w\_final, and CONTROL files from ShengBTE. "DATABASE" is used to directly read the phonon properties from the GiftBTE database. It provides the heat capacity $C$, relaxation time $\tau$, and group velocity $\bf{v}$ of convergent bands for some materials. After defining the type, the path of the phonon properties files needs to be provided.
\item \textbf{GeometryMaterialType}: This command defines the corresponding material for each geometry region in the mesh file (Geometry module). 
\item \textbf{MaterialDimension}: This command defines the material dimension. "2" is used for two-dimensional materials such as graphene. "3" is used for three-dimensional materials like Si, GaN, et al. Note that the material dimension does not necessarily correlate with the geometry dimension. There is also one-dimensional geometry for three-dimensional materials (silicon thin film). 
\item \textbf{Ntheta}: This command defines the number of polar angles $\theta$ using directional discretization.
\item \textbf{Nphi}: This command defines the number of azimuth angles $\phi$ using directional discretization.
\end{itemize}

\subsection{Result}
In the Results section, the temperature and heat flux values are derived from the energy density and subsequently displayed. Several output files are generated, including:

\begin{itemize}
\item \textbf{Temperature.dat}: This file contains the temperature $T$ at the center of each cell.
\item \textbf{TemperatureLattice.dat}: This file contains the lattice temperature $T_L$ at the center of each cell.
\item \textbf{HeatFlux.dat}: This file contains the heat flux vector $\bf{q}$ at the center of each cell.
\item \textbf{Result.vtk}: This file contains temperature and heat flux values in the vtu format, which can be visualized using Paraview (www.paraview.org) for graphical representation.

\end{itemize}

\section{Investigation on submicron thermal transport} \label{sec:investigation}
As such, the application of GiftBTE encompasses various tasks, including but not limited to computing the thermal conductivity of nanostructures, predicting temperature rises in transistors, and simulating laser heating processes (processes with small hot spots or ultra-fast processes). In this section, we employ GiftBTE to investigate these three types of problems, using Si as the material of interest. All of these cases have been listed as examples in the GiftBTE package. The phonon properties of Si are obtained through first-principles calculations, employing the anharmonic lattice dynamics method implemented in the ALAMODE package \citep{tadano2014anharmonic}. Our calculations employ a 50×50×50 grid of q-points to sample the Brillouin zone, resulting in 50×50×50×6 phonon modes. The calculated bulk thermal conductivity of Si is found to be 145 W/m-K, which is in excellent agreement with the literature values \citep{lindsay2018survey,tadano2014anharmonic}. The phonon properties are then taken as the input of the phonon module.
\subsection{Thermal conductivity of nano-porous media}
\begin{figure}[!htb]
    	\centering
    	\includegraphics[width=1\textwidth]{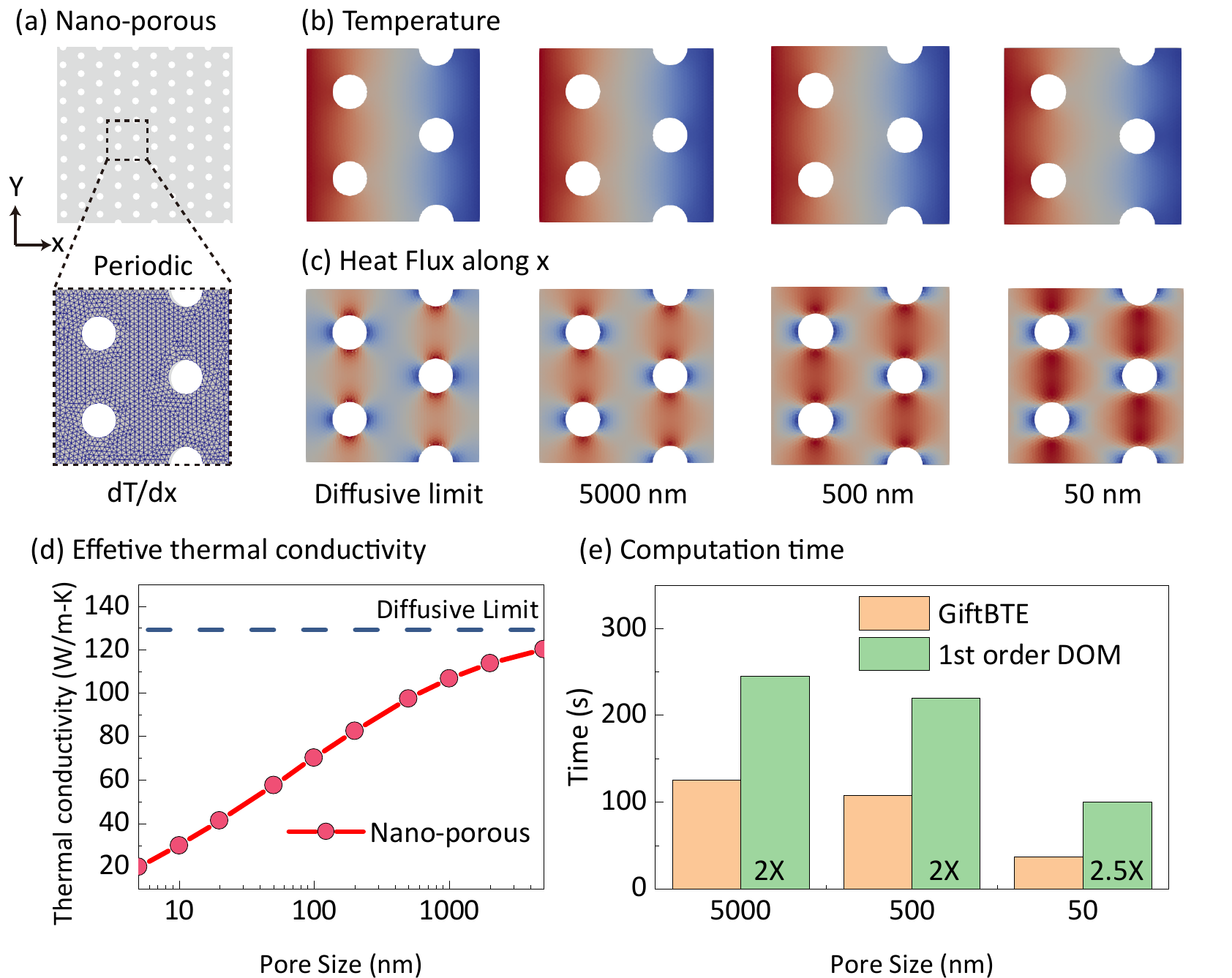}
    	\caption{\textbf{Thermal conductivity of nano-porous media} (a) The structure and the computational domain of the nano-porous media. (b) Temperature profiles of nano-porous media. (c) Heat flux profiles ($x$ direction) of nano-porous media. (d) Effective thermal conductivity of nano-porous media. (e) Computation time of GiftBTE and implicit DOM with first-order spatial accuracy.}
    	\label{fig:porous}
\end{figure}

We illustrate the application of GiftBTE by considering nano-porous media as a representative example of nanostructures and calculating their effective thermal conductivity. The structure and computational domain of the nano-porous media are depicted in Figure \ref{fig:porous}(a), while the porosity of the material is determined to be 15.7\%. To facilitate computation, we extract a single period of the nano-porous media for analysis, employing periodic boundary conditions on all outer boundaries. Additionally, a temperature difference of 1 K is imposed between the left and right boundaries, resulting in a temperature gradient along the $x$ direction. The surface of the pore is treated as a diffuse boundary \citep{wei2022quantifying}. Triangular meshes are employed to discretize the computational domain, as depicted in Figure \ref{fig:porous}(a). We perform computations for various pore sizes, ensuring convergence through a convergence test. For these computations, we employ 12 phonon bands, 512 directions, and 2900 meshes. Considering the rapid convergence observed in all cases using the sequential iterative method, we utilize the sequential iterative method within GiftBTE.

The temperature profiles and heat flux profiles (along the $x$ direction) for various scales are illustrated in Figure \ref{fig:porous}(b) and (c), respectively. It is evident that despite similar temperature profiles, the heat flux profile significantly differs for varying pore sizes. The variation in effective thermal conductivity with respect to pore size is depicted in Figure \ref{fig:porous}(d). A finite size effect on thermal conductivity (smaller than the effective thermal conductivity predicted by the macroscopic heat diffusion equation, which is denoted as "diffusive limit") is observed for pore sizes less than 5 $\upmu$m.

The remarkable efficiency of the GiftBTE is demonstrated by presenting a summary of the computational times for the aforementioned cases, as illustrated in Fig. \ref{fig:porous}(e). The computational performance evaluation is conducted on a cluster equipped with Intel Xeon ICX Platinum 8358 processors, utilizing 4 CPU cores. Notably, all simulations were completed within 2 minutes.

To further showcase the exceptional performance of GiftBTE, a comparison of computational time is made with the first-order implicit DOM (the implicit DOM with first-order accuracy and sequential iterative method) employed in the previous package \citep{romano2021openbte}. After conducting convergence tests, a total of 9600 triangular meshes are utilized for simulating with the first-order DOM. In comparison, the current solver reduces the computational cost by a factor of 2.5. This acceleration is achieved due to the fact that first-order accuracy necessitates a larger number of meshes.

\subsection{Temperature rise of fin field-effect transistor}
\begin{figure}[!htb]
    	\centering
    	\includegraphics[width=1\textwidth]{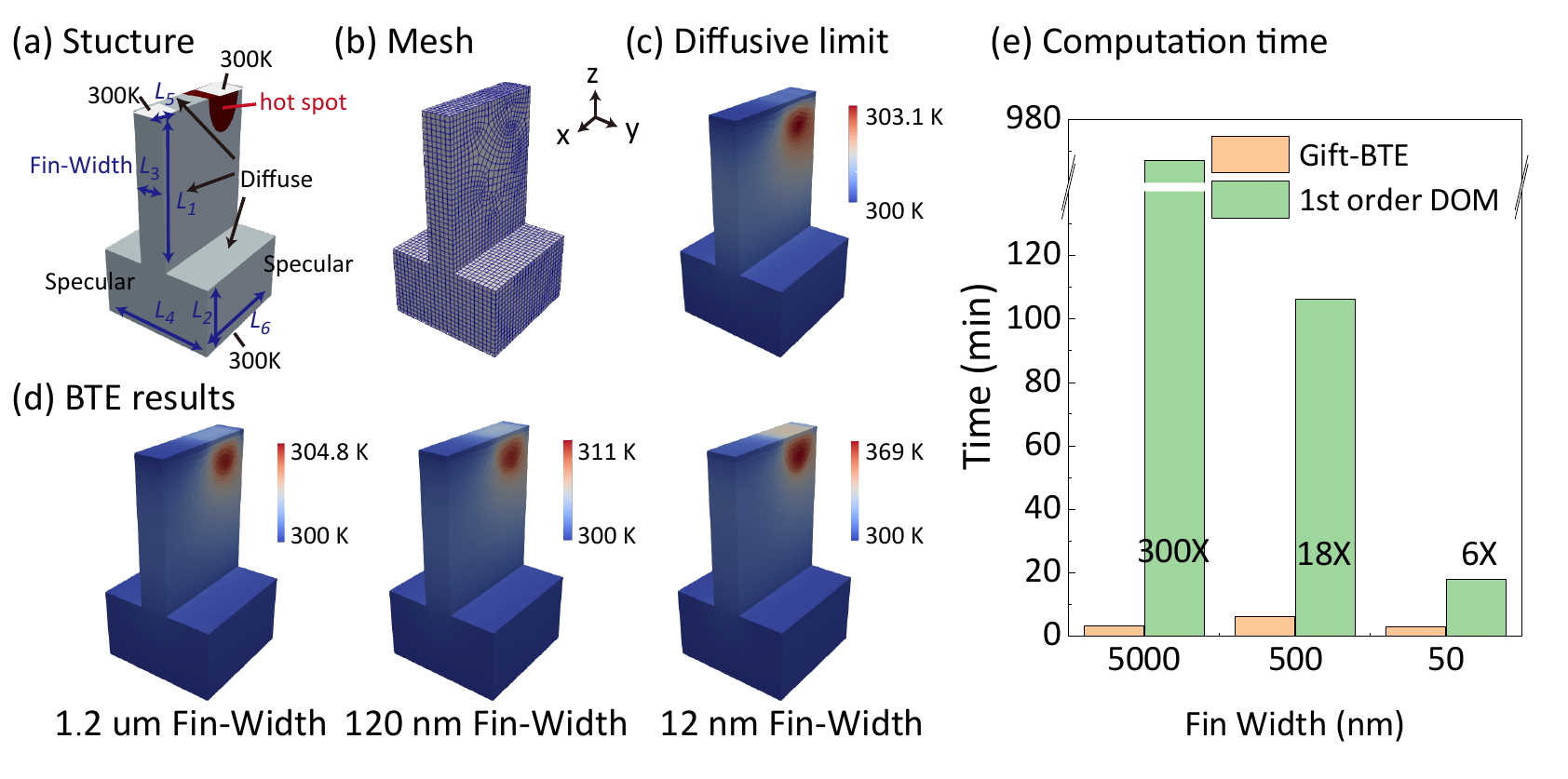}
    	\caption{\textbf{Temperature field of FinFET} (a) The structure and the computational domain of the FinFET. The red part is the hot spot. (b) Unstructured mesh to discrete the FinFET. (c) Temperature field of FinFET predicted by the heat diffusion equation. (d) Temperature field of FinFET predicted by the GiftBTE. (e) Computation time of GiftBTE and first-order implicit DOM.}.
    	\label{fig:FinFET}
\end{figure}

We select the fin field-effect transistor (FinFET) as a representative example of transistors and employed GiftBTE to predict its temperature distribution. The structure and boundary conditions of the FinFET are illustrated in Figure \ref{fig:FinFET}(a), which are adapted from a previous study \citep{wang2015electron}. The thermalizing boundary, in contact with the metal electrode (source and drain) or away from the hot spot (substrate) \citep{hao2018electrothermal,majumdar1993microscale}, is set at 300 K to simulate a room temperature environment \citep{maiti2017introducing}. The diffusely reflecting boundary is set at the boundary in contact with the dielectric layer \citep{wang2015electron}. The other boundaries are set as specularly reflecting boundaries to mimic the symmetric boundary between devices \citep{hu2020unification,sheng2022size}. The hot spot is set as a semi-ellipse near the drain and is commonly found in enhancement transistors \citep{hao2018hybrid}. The detailed dimensionless geometrical parameters are presented in Table \ref{tab:table_1}. The absolute values for these parameters are tested at different scales by multiplying the ratio scales of 1 nm, 10 nm, and 100 nm (i.e., the cases have a fin-width of 12 nm, 120 nm, and 1.2 $\upmu$m) to check the performance of this solver from the ballistic regime to the diffusive regime. The total volumetric heat generation rates $\sum\limits_{\lambda} {{{\dot Q}_{\lambda}}} $ inside the hot spot are ${10^{19}}$ W/m$^3$,${10^{17}}$ W/m$^3$,${10^{15}}$ W/m$^3$ for the fin-widths of 12 nm, 120 nm, and 1.2 $\upmu$m, respectively. These volumetric heat generation rates are chosen to ensure that the temperature profiles for different cases are the same if they are solved by the heat diffusion equation, which adopts the bulk thermal conductivity of Si: 145 W/m-K (the result is shown in Fig. \ref{fig:FinFET}(c)). The mode-level heat generation rates are assumed to be uniformly distributed in different directions and proportional to the heat capacity for different bands, mimicking an equilibrium heat source \citep{hu2020unification}. The properties of the heat generation in this study are set for demonstration purposes. To capture the rigorous heat generation due to phonons emitted by hot electrons, the electron BTE considering electron-phonon coupling needs to be solved simultaneously \citep{hao2018electrothermal, pop2010energy, hao2017hybrid, ashok2010electrothermal}. After conducting a convergence test, we utilized 12 phonon bands, 128 directions, and 14,442 hexahedron meshes for our analysis. As an illustrative example, Fig. \ref{fig:FinFET}(b) depicts the unstructured hexahedron meshes employed for discretizing the FinFET.

The temperature profiles for various scales are presented in Figure \ref{fig:FinFET}(d) using GiftBTE. Due to the exceedingly slow convergence rate observed in the 1.2 $\upmu$m case, the synthetic iterative scheme is employed in the computation of FinFETs. As the size decreases, the disparity between the results obtained from the BTE computation and the heat diffusion computation becomes more pronounced. For a 12 nm fin-width, the BTE predicts a maximum temperature increase of 85 K, which is approximately 27 times higher than the temperature rise predicted by the heat diffusion equation (Figure \ref{fig:FinFET}(c)). This observation signifies the non-Fourier effects that arise due to ballistic transport and boundary scattering, particularly in small-sized devices \citep{chen2021non}. Higher temperature rises, as revealed by the BTE simulations, impose more severe limitations on the electrical performance of nanoscale devices \citep{rhyner2016minimizing}, underscoring the importance of comprehending the self-heating effect. The aforementioned results demonstrate the capability of GiftBTE to perform full three-dimensional nanoscale simulations spanning the ballistic to diffusive regimes.

\begin{table}[]
    \centering
    \setlength{\abovecaptionskip}{10pt}
    \setlength{\belowcaptionskip}{10pt}
    \caption{The dimensionless geometrical parameters of the FinFET.}
    \setlength{\tabcolsep}{7mm}{
    \begin{tabular}{cccccccc}
    \hline
        $L_1$ & $L_2$ & $L_3$ & $L_4$ & $L_5$ & $L_6$  \\
    \hline
         $60$ & $30$ & $12$ & $50$ & $12.5$ & $50$ \\
    \hline
        
    \end{tabular}
    }
    \label{tab:table_1}
\end{table}

The outstanding efficiency of the current solver is showcased through the summarized computational times of the aforementioned cases, as illustrated in Figure \ref{fig:FinFET}(e). The computational performance evaluation is conducted on a cluster of Intel Xeon ICX Platinum 8358 processes, utilizing 16 CPU cores. Remarkably, all simulations were completed within 5 minutes.
 
To further demonstrate the exceptional performance of the present solver, a comparison is made between the computational time of the present solver and the first-order implicit DOM (implicit DOM with first-order spatial accuracy and sequential iterative scheme), which was utilized in the previous package \citep{romano2021openbte}. After conducting convergence tests, a simulation is performed using 110,000 hexahedral meshes with the first-order DOM. The present solver significantly reduces the computational cost by a factor of between 6 and 300. This acceleration is attributed to two factors: firstly, the first-order accuracy necessitates a larger number of meshes; secondly, the synthetic iterative scheme substantially reduces the number of iterations as compared to the sequential iterative scheme in present cases.

\subsection{Time-domain thermoreflectance experiment}

\begin{figure}[!htb]  
\centering  
\includegraphics[width=1\textwidth]{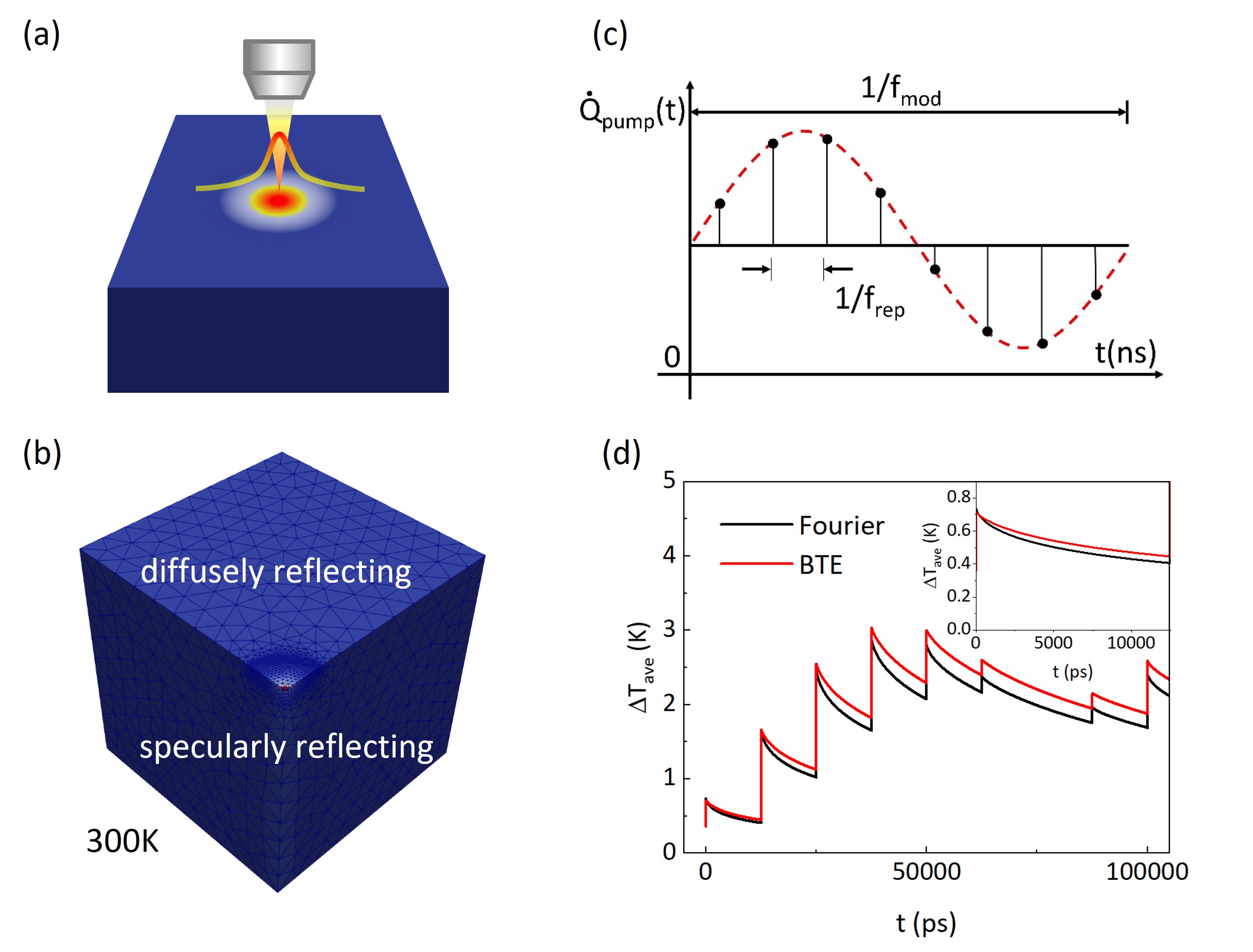}
  \caption{\textbf{Time-domain thermoreflectance Simulation} (a) Pump in TDTR technique (b) computational domain and boundary conditions. (c) The modulated power pulse. (d) Comparison of the temperature rise obtained using the phonon BTE and the heat diffusion equation (Fourier).}
\label{fig:TDTR}  

\end{figure}

We select the time-domain thermoreflectance experiment (TDTR) as an example of the laser heating process. The structure of the TDTR experiment is depicted in Figure \ref{fig:TDTR}(a), where an ultra-fast modulated pulsed laser directly illuminates the semiconductor \citep{wang2016thermal}. The computational domain and set of boundary conditions are illustrated in Figure \ref{fig:TDTR}(b). Due to symmetry, we simulated a quarter of the entire system, as each side of the cube is 50 $\mu$m. The top boundary represents the surface of the semiconductor and is set as a diffusely reflecting boundary. The two inner boundaries are set as specularly reflecting boundaries to capture the symmetry. The remaining three boundaries are set at 300 K to simulate a room temperature environment. At the initial state, the entire computational domain is at the equilibrium state of 300 K.

The heat generation induced by the pump beam in the sample can be described as follows \cite{jiang2018tutorial}:

\begin{equation}
\dot{Q}_{\text{pump}}(r,z,t) = \dot{Q}_0 e^{-2r^2/w_0^2} e^{- z/h} (1 + \sin(tf_{\text{mod}})) \sum_{n=-\infty}^{\infty} \delta(t - n/f_{\text{rep}}).
\end{equation}

Here, $\dot{Q}_0$ represents the average power of the pump beam, set to $4.64\times18  $ W/m$^3$. $r$ is the distance between each position and the center of the pump beam, while $w_0$ represents the radius of the pump beam, set as 3.8 $\mu$m. The intensity of the pump decays exponentially with a penetration depth $h = 1.5 \mu$m \citep{TDTR1}. The heat generation is distributed across each phonon mode according to its heat capacity. Each pulse has a duration of 1 ps, short enough to be expressed as the delta function $\sum_{n=-\infty}^{\infty} \delta(t - n/f_{\text{rep}})$ in the equation. The repetition frequency of each pulse $f_{\text{rep}}$ is set as 80 MHz. $f_{\text{mod}}$ represents the modulation frequency, set as 10 MHz. Figure \ref{fig:TDTR}(c) displays the power variation at different times for a given repetition frequency and modulation frequency.

After conducting a convergence test, we employed 28,595 tetrahedral meshes, 32 directions, and 10 phonon bands to investigate this problem. Figure \ref{fig:TDTR}(d) illustrates the average temperature at the top boundary, $T_{\text{ave}}$, as it changes over time within a single modulation period. $T_{\text{ave}}$ is expressed as:

\begin{equation}
T_{\text{ave}} = \frac{{\sum_{r<2w_0} T(r,t) e^{-2r^2/w_0^2}}}{{\sum_{r<2w_0} e^{-2r^2/w_0^2}}}
\end{equation}

The results obtained from solving the phonon BTE demonstrate a slower temperature decay compared to that obtained from the heat diffusion equation (Fourier). The computation is conducted on a cluster of Intel Xeon ICX Platinum 8358 with 64 CPU cores and took 16 hours.

\section{Summary and conclusions} \label{sec:conclu}

In this study, we demonstrate an efficient deterministic solver, GiftBTE, for the phonon BTE that is applicable to arbitrary crystalline materials and devices. This solver incorporates both steady-state and transient solvers. The steady-state solver employs the implicit DOM with second-order spatial accuracy and a synthetic iterative scheme, while the transient solver utilizes the explicit DOM with second-order spatial accuracy. By interfacing with first-principles calculations, this solver enables computation of submicron thermal transport for arbitary crystalline materials. With GiftBTE, three types of problems can be investigated: (i) computation of thermal conductivity for nanostructures, (ii) prediction of temperature rise in transistors, and (iii) simulation of laser heating processes. In this study, we explore these three types of problems using GiftBTE and demonstrate its excellent efficiency. In these problems, significant non-Fourier effects are observed due to the existence of submicron sized local structures or submicron sized heat sources. The GiftBTE package and its manual can be accessed through the website: https://github.com/GiftBTE-developer/GiftBTE.

\section*{Acknowledgments}
This work is supported by the National Key R \& D Project from Ministery of Science and Technology of China (Grant No. 2022YFA1203100) and the National Natural Science Foundation of China (Grant No. 52122606). The computations in this paper were run on the $\pi$ 2.0 cluster cluster supported by the Center for High Performance Computing at Shanghai Jiao Tong University.


 \bibliographystyle{elsarticle-num}





\end{document}